\documentclass[aps,groupedaddress,twocolumn,floatfix,showpacs,tightenlines]{revtex4-1}
\usepackage{graphicx}
\usepackage{float}
\usepackage{color}
\usepackage{psfrag}
\usepackage{dcolumn}% Align table columns on decimal point
\usepackage{bm}% bold math

\usepackage{amssymb}
\usepackage{amsmath}
\usepackage{amsfonts}
\usepackage{longtable}
\usepackage{xspace}
\usepackage{mathrsfs}
\usepackage{gensymb}

\newcommand{\beq}{\begin{equation}}
\newcommand{\eeq}{\end{equation}}
\newcommand{\bea}{\begin{eqnarray}}
\newcommand{\eea}{\end{eqnarray}}
\newcommand{\bes}{\begin{subequations}}
\newcommand{\ees}{\end{subequations}}

\newcommand{\scri}{\mathscr{I}}

\begin{document}
\title{Adapted gauge to a quasilocal measure of the black holes recoil}
\author{James Healy, Carlos O. Lousto, Nicole Rosato}
\affiliation{Center for Computational Relativity and Gravitation,\\
School of Mathematical Sciences,
Rochester Institute of Technology, 85 Lomb Memorial Drive, Rochester,
New York 14623}
%\date{7/24/2019}
\date{\today}

\begin{abstract}
We explore different gauge choices in the moving puncture formulation
in order to improve the accuracy of a linear momentum measure evaluated
on the horizon of the remnant black hole produced by the merger of a binary.
In particular, motivated by the study of gauges in which $m\eta$
takes on a constant value, 
we design a gauge via a variable shift parameter $m\eta(\vec{r}(t))$. 
This new parameter takes a low value asymptotically, as well as at the
locations of the orbiting punctures,
and then takes on a value of approximately 2 at the final hole horizon.
This choice then follows the remnant black hole as it moves due to its
net recoil velocity.
We find that this choice keeps the accuracy of the binary evolution. 
Furthermore, if the asymptotic value of the parameter $m\eta$ is chosen about or
below 1.0, it produces more accurate results for the recoil velocity than the
corresponding evaluation of the radiated linear momentum at infinity,
for typical numerical resolutions.
%We also find that the choice of the
%$\partial_t$-gauge (at our working resolutions)
%is more accurate in this regard of computing recoil velocities than the
%$\partial_0$-gauge. 
Detailed studies of an unequal mass
$q=m_1/m_2=1/3$ nonspinning binary are provided and then verified for
other mass ratios $(q=1/2,1/5)$ and spinning $(q=1)$ binary black hole mergers.
\end{abstract}
\pacs{04.25.dg, 04.25.Nx, 04.30.Db, 04.70.Bw}

\maketitle

%%%%%%%

\section{Introduction}\label{sec:intro}

The discovery by numerical relativity computations 
\cite{Campanelli:2007ew,Gonzalez:2007hi,Campanelli:2007cga}
that binary black hole mergers may impart thousand
of kilometers per second speeds to the final black hole remnant
had an immediate impact on the interest of observational 
astrophysics to search for signatures of such recoil for
supermassive black holes in merged galaxies 
(See \cite{Komossa:2012cy}, for an early review).
The interest in searching for observational effects extends
to nowadays \cite{Chiaberge:2016eqf,Lousto:2017uav,Chiaberge:2018lkg}, 
including its incidence in statistical distributions 
\cite{Lousto:2009ka,Fishbach:2017dwv,Lousto:2012su} and binary formation
channels as well as cosmological consequences \cite{Rodriguez:2019huv}.

More recently theoretical explorations evaluate the possibility
to directly detect the effects of recoil on the gravitational
waves observed by LIGO \cite{Lousto:2019lyf} and 
LISA \cite{Sesana:2008ur,Gerosa:2016vip}.
The use of numerical relativity waveforms to directly compare
with the observation of gravitational waves require accurate 
modeling and good coverage of the parameter space
\cite{Abbott:2016apu}. There are
already successful descriptions for the 
GW150914 \cite{Lovelace:2016uwp,Healy:2019jyf} 
and GW170104\cite{Heal:2017abq} 
events and the analysis of the rest of the 
O1/O2 events \cite{LIGOScientific:2018mvr}
is well underway (Healy et al. 2020a, paper in preparation).

The accurate modeling of the final remnant of the merger of binary 
black holes is also of high interest for applications to gravitational
waves modeling and tests of gravity as a consistency check
\cite{TheLIGOScientific:2016src,LIGOScientific:2019fpa}.
The computation of the final remnant mass and spin can be performed
in three independent ways, a fit to the quasinormal modes of the 
final remnant Kerr black hole \cite{Echeverria89,Berti:2005ys,Berti:2009kk},
a computation of the energy and angular momentum carried away
by the gravitational radiation to evaluate the deficit from the initial
to final mass and spins, and a quasilocal computation of the horizon
mass and spin using the isolated horizon formulas \cite{Dreyer:2002mx}.
Comparison of the three methods has been carried out in
\cite{Dain:2008ck,Lousto:2019lyf}, concluding that at the typical resolutions
used in production numerical relativity simulations the horizon
quasilocal measures are an order of magnitude more accurate than
the radiation or quasinormal modes fittings.

This leads to very accurate modeling of the final mass and spins
from their initial binary parameters. In particular for
nonprecessing binaries, the modeling 
\cite{Healy:2018swt} warrants errors typically $0.03\%$ for the mass and $0.16\%$ for 
the spin.
Meanwhile the modeling of the final recoils leads to errors of
the order of $5\%$ since radiation of linear momentum is used
to evaluate them.
A similar accurate modeling of the recoil could be attempted
by the use of a horizon quasilocal measure.

In reference \cite{Krishnan:2007pu} a quasi-local formula
for the linear momentum of black-hole horizons was proposed,
inspired by the
formalism of quasi-local horizons. This formula was tested using two
complementary configurations: (i) by calculating the large orbital
linear momentum of the two black holes in an orbiting, unequal-mass,
zero-spin, quasi-circular binary and (ii) by calculating the very
small recoil momentum imparted to the remnant of the head-on collision
of an equal-mass, anti-aligned-spin binary. The results obtained were
consistent with the horizon trajectory in the orbiting case, and
consistent with the radiated linear momentum for the much smaller
head-on recoil velocity. A key observation we will explore in this
paper is the dependence of the accuracy on a gauge parameter
used in our simulations.

This paper is organized as follows,
In Section \ref{sec:nr} we study in detail the effects of choosing
different (constant) values of $\eta$, the damping parameter in the shift
evolution equation on the accuracy of the quasilocal measure of the
horizon linear momentum proposed in \cite{Krishnan:2007pu}.
We discuss in detail a prototype case of a nonspinning $q=1/3$ binary.
Other unequal mass cases and one spinning case are verified as well.
In Section \ref{sec:OtherStudies}, we perform additional studies of
the shift evolution equation for the alternative $\partial_0$-gauge,
variable $\eta$, and apply what we learned in the previous section
to develop a variable shift parameter $\eta$ and to more extreme
unequal mass binary black hole mergers.
In Section \ref{sec:dis} we discuss the benefits of the using of different
values of $\eta$ from our standard $\eta=2$
(See also \cite{vanMeter:2006vi}) for generic simulations,
in particular for those that involve an accurate computation of the
remnant recoil and we also conclude by noting the advantage of keeping
the $\partial_t$-gauge for evolutions over the $\partial_0$-gauge.

%%%%%%%%%%%%
\section{Numerical Techniques}\label{sec:nr}
%%%%%%%%%%%%

Since the 2005 breakthrough work \cite{Campanelli:2005dd}
We obtain accurate, convergent waveforms and horizon parameters by
evolving the BSSNOK \cite{Nakamura87, Shibata95, Baumgarte99}
system in conjunction with a modified 1+log lapse and a
modified Gamma-driver shift condition~\cite{Alcubierre02a,Campanelli:2005dd},
  \begin{eqnarray}
\partial_0\alpha=(\partial_t - \beta^i \partial_i) \alpha &=& - 2 \alpha K,\\
 \partial_t \beta^a &=& \frac34 \tilde \Gamma^a - \eta(x^k,t)\, \beta^a.\label{eq:gauge}
  \end{eqnarray}
with an initial vanishing shift and lapse $\alpha(t=0) = 2/(1+\psi_{BL}^{4})$. 
Here, and for the remainder of the paper, latin indices cover the spatial range $i=1,\cdots,3$.

An alternative moving puncture evolution can be achieved \cite{Baker:2005vv}
by choosing \cite{vanMeter:2006vi}
  \begin{eqnarray}
\partial_0\alpha=(\partial_t - \beta^i \partial_i) \alpha &=& - 2 \alpha K,\\
\partial_0 \beta^a=(\partial_t - \beta^i \partial_i) \beta^a &=& \frac34 \tilde \Gamma^a - \eta(x^k,t)\, \beta^a.\label{eq:gaugenasa}
  \end{eqnarray}
In the subsequent, we will refer to this first order equations for the shift (\ref{eq:gauge}) as the $\partial_t$-gauge and to (\ref{eq:gaugenasa}) as the $\partial_0$-gauge. Unless otherwise stated, all binary black hole simulations in this paper use the $\partial_t$ gauge.

The parameter $\eta$ (with dimension of one-over-mass: $1/m$) in the shift equation regulates the damping of the gauge oscillations and is commonly chosen to be of order unity (we use $\eta=2/m$) as a compromise between the accuracy and stability of binary black hole evolutions.
We have found in \cite{Lousto:2007db} that coordinate dependent measurements, such as spin and linear momentum direction, become more accurate as $\eta$ is reduced (and resolution $h\to0$). However, if $\eta$ is too small $(\eta\ll1/m)$, the runs may become unstable. Similarly, if $\eta$ is too large $(\eta\gg10/m)$, then grid stretching effects can cause the remnant horizon to continuously grow, eventually leading to an unacceptable loss in accuracy at late times.

We use the {\sc TwoPunctures}~\cite{Ansorg:2004ds} thorn to compute initial data.
We evolve these black-hole-binary data-sets using the
{\sc LazEv}~\cite{Zlochower:2005bj} implementation of the moving
puncture formalism~\cite{Campanelli:2005dd}.
We use the Carpet~\cite{Schnetter-etal-03b,carpet_web} mesh refinement
driver to provide a `moving boxes' style mesh refinement and
we use {\sc AHFinderDirect}~\cite{Thornburg2003:AH-finding} to locate
apparent horizons.
We compute the magnitude of the horizon spin using
the {\it isolated horizon} (IH) algorithm detailed in
Ref.~\cite{Dreyer02a} (as  implemented in
Ref.~\cite{Campanelli:2006fy}).
Once we have the horizon spin, we can calculate the horizon
mass via the Christodoulou formula 
${m_H} = \sqrt{m_{\rm irr}^2 + S_H^2/(4 m_{\rm irr}^2)}\,,$
where $m_{\rm irr} = \sqrt{A/(16 \pi)}$ and  $A$ is the surface area
of the horizon. 
We measure radiated energy, linear momentum, and angular momentum, in
terms of $\psi_4$, using the formulae provided in
Refs.~\cite{Campanelli99,Lousto:2007mh} and extrapolation to to $\scri^+$
is performed with the formulas given in Ref.~\cite{Nakano:2015pta}.

Convergence studies of our simulations have been performed
in Appendix A of Ref.~\cite{Healy:2014yta},
in Appendix B of Ref.~\cite{Healy:2016lce}, and
for nonspinning binaries are reported in Ref.~\cite{Healy:2017mvh}.
For very highly spinning black holes ($s/m^2=0.99$)
convergence of evolutions was studied in Ref. \cite{Zlochower:2017bbg},
for precessing $s/m^2=0.97$ in Ref. \cite{Lousto:2019lyf}, and
for ($s/m^2=0.95$) in Ref. \cite{Healy:2017vuz} for unequal mass binaries.
These studies allow us to assess that the simulations presented below,
with similar grid structures, are well
resolved by the adopted resolutions and are in a convergence regime.

In Reference ~\cite{Krishnan:2007pu}
we introduced an alternative quasi-local measurement of the
linear momentum of the individual (and final) black holes in the binary
that is based on the coordinate rotation and translation vectors
\begin{equation}
  P_{[i]} = \frac{1}{8\pi}\oint_{AH} \xi^a_{[i]} R^b (K_{ab} - K \gamma_{ab}) d^2V,
  \label{eq:coordmom}
\end{equation}
where $K_{ab}$ is the extrinsic curvature of the 3D-slice, $d^2V$ is the
natural volume element intrinsic to the horizon, $R^a$ is the
outward pointing unit vector normal to the horizon on the 3D-slice,
and $\xi^i_{[\ell]} = \delta^i_\ell$.

We tested this formula using two complementary configurations: (i) by
calculating the large orbital linear momentum of the two unequal-mass
$(q=1/3)$, nonspinning, black holes in a quasi-circular orbit and (ii)
by calculating the very small recoil momentum imparted to the remnant
of the head-on collision of an equal-mass, anti-aligned-spin binary.
When we reduce the gauge parameter $m\eta$ from 2 to 1 in the orbiting case,
we obtain results consistent with the horizon trajectory.
Similarly for the head-on case we find results consistent with the net radiated
linear momentum, however the remainder of the paper will focus on only 
the orbiting case.
%We obtain results consistent with the horizon trajectory in the
%orbiting case, and consistent with the net radiated linear momentum
%for the much smaller head-on recoil velocity when we reduced the gauge
%parameter from $m\eta=2$ to $m\eta=1$.

Here we explore this initial results in much more detail, allowing
for even smaller values of $\eta$ and assessing convergence of the
results with both, numerical resolution and values of $\eta\to0$.
This will allow us to assess when the quasilocal measure of linear
momentum (\ref{eq:coordmom}) can be considered more accurate than
the measure of radiated linear momentum at $\scri^+$.

In our simulations, we normalize data such that the sum of the horizon
masses, after spurious radiation of initial data, is set to unity, i.e. 
$m_{H1}+m_{H2}=1$. In the tables below, we also introduce the difference of
the ADM mass and angular momentum minus the final black hole mass and spins,
as $\Delta m=M_{ADM}-m_{f}$ and  $\Delta J=J_{ADM}-\alpha_{f}$.

%%%%%%%%%%%%
\subsection{Results for a $q=1/3$ nonspinning binary}\label{sec:results}
%%%%%%%%%%%%

As a prototypical case of study we will consider a binary with mass
ratio $q=m_1/m_2=1/3$ and spinless black holes starting at an initial
coordinate separation $D=9m$, with $m=m_1+m_2$ the total mass of the
system. From this separation the binary performs about 6 orbits
before the merger into a single final black hole at
around $t=725m$.

\begin{figure}[h]
  \includegraphics[angle=0,width=0.9\columnwidth]{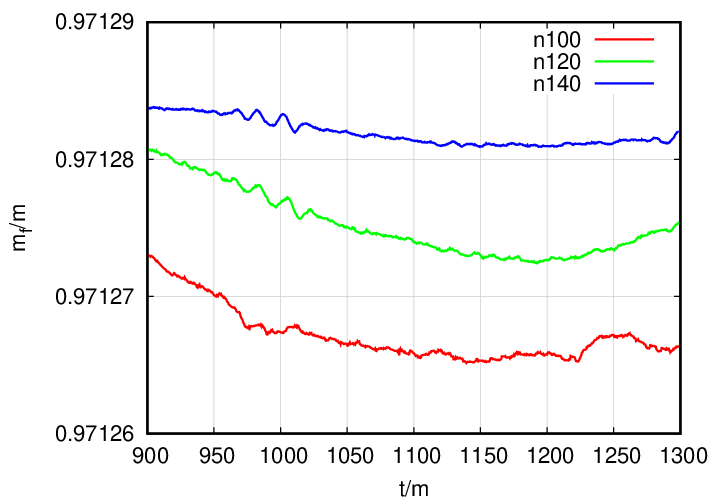}
      \includegraphics[angle=0,width=0.9\columnwidth]{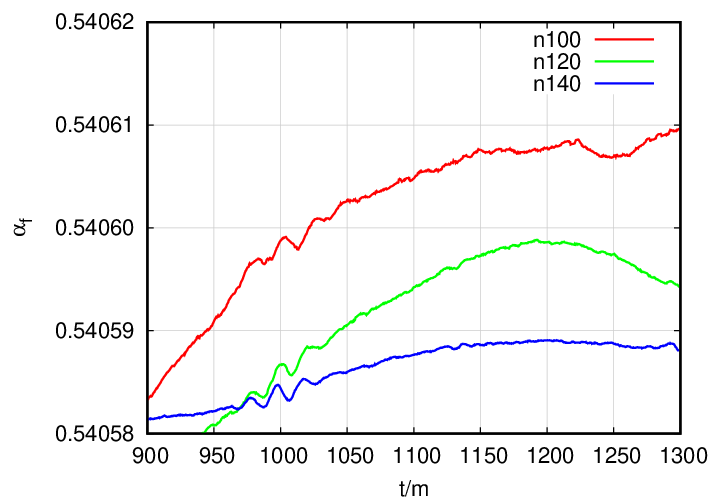}
      \caption{The horizon measure of the mass and spin after merger of a $q=1/3$ nonspinning binary versus time for $m\eta=2$ at resolutions n100, n120, n140.
    \label{fig:mfaf}}
\end{figure}

The final mass and final spin are measured very
accurately by the horizon quasilocal formulas
\cite{Dreyer:2002mx,Campanelli:2006fy}; figs.~\ref{fig:mfaf}
provide a visualization of their respective values after merger
into the final settling black hole remnant.
They display smaller variations versus time with increasing
resolutions.

\begin{figure}[h]
  \includegraphics[angle=0,width=\columnwidth]{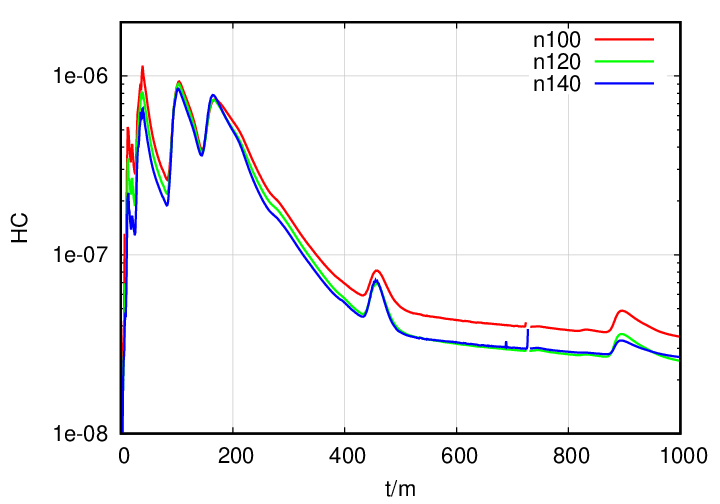}
  \includegraphics[angle=0,width=0.9\columnwidth]{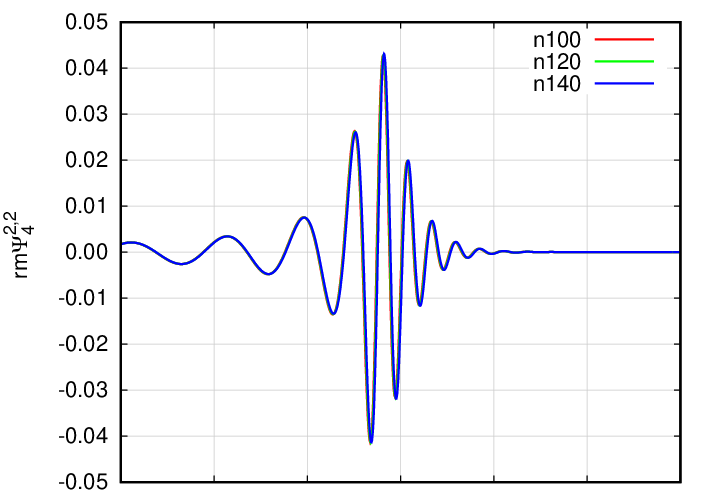}
      \includegraphics[angle=0,width=0.9\columnwidth]{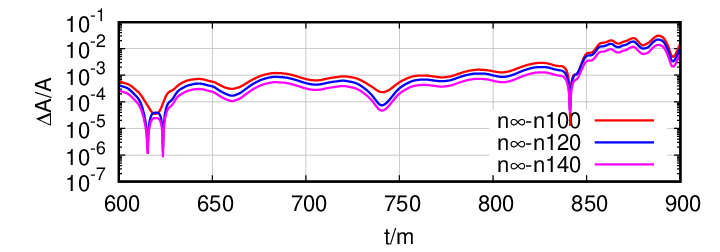}
      \includegraphics[angle=0,width=0.9\columnwidth]{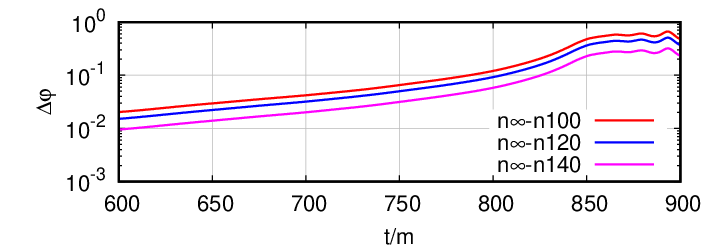}
      \caption{The Hamiltonian constraint behavior versus time for $m\eta=2$ at resolutions n100, n120, n140 in the top panel, and in the bottom panels, the (2,2)-waveform as seen by an observer at $R=113m$ for the $q=1/3$ nonspinning binary from an initial separation $D=9m$, the amplitude difference between resolutions, $\Delta A/A$, and the phase difference (in radians), $\Delta \varphi$, with respect to the infinite resolution extrapolation $(n_\infty)$. \label{fig:HCWF}}
\end{figure}

In Figs.~\ref{fig:HCWF}
the convergence of the Hamiltonian constraint
(Momentum constraints show a very similar convergent behavior)
and the merger gravitational waveforms $(\ell,m)=(2,2)$-mode
are displayed. Both show highly
convergent behavior, therefore they are
%well in the numerical convergence regime and
resolving the binary system accurately.

As shown in the Tables \ref{table:I} and \ref{table:II} the
computed final mass and spin of the remnant black hole are
well in the convergence regime at typical computation
resolutions \cite{Healy:2017psd,Healy:2019jyf}.

%\begin{widetext}
\begin{table}[h!]
  \caption{The final black hole mass ($m_f/m$) for different $\eta$ and resolutions
(top table).
Difference between final black hole and ADM masses (bottom table). 
Values for $m\eta\to0.0$ and resolution $h_i\to0$ are extrapolated
(the later by order $n$).}
\begin{ruledtabular} 
  \begin{tabular}{lllll}
$h_i/m$&$m\eta=2.0$&$m\eta=1.0$&$m\eta=0.5$&$m\eta\to0.0$\\%&order\\
%$h_i/m\backslash m\eta$ & $2.0$ & $1.0$&$0.5$&$\to0.0$\\%&order\\
\hline
1/100&0.97127	&0.97127	&0.97128	&0.97127	\\
1/120&0.97128	&0.97129	&0.97126	&0.97128	\\
1/140&0.97128	&0.97128	&0.97128	&0.97128	\\
$\to0$ &0.97131 &0.97135 &0.97135 &0.97132\\
$n$ &1.59 &2.24 &1.95 &2.29\\
\hline
1/100&0.02046&0.02046&0.02047&0.02046\\%&-2.49\\
1/120&0.02046&0.02045&0.02045&0.02044\\%&0.28\\
1/140&0.02045&0.02045&0.02045&0.02045\\%&-2.40\\
%$\to0$&0.0205071&0.0000000&0.0000000& \\
%order   &-0.2761455&0.0000000&0.0000000& \\
  \end{tabular} \label{table:I}  
\end{ruledtabular}
\end{table}
%\end{widetext}

Table~\ref{table:I} displays the 
difference between the initial total ADM mass and
final horizon mass and 
Table~\ref{table:II} displays the loss of ADM angular momentum
from its initial value to the
final horizon spin
for different resolutions and different
($\eta$-values) gauges. The computations give consistent values
to 5-decimal places for each resolution, showing we are deep in
the convergence regime and also versus $\eta$, showing (as expected)
that those computations are gauge-invariant.

%\begin{widetext}
\begin{table}[h] 
  \caption{The final black hole spin ($\alpha_f/m^2$) for different $\eta$ and resolutions
(top table).
Difference between final black hole angular momentum $\alpha_f$ and initial ADM angular momentum $J_{ADM}$
(bottom table). Values for $m\eta=0.0$ are extrapolated as those for
infinite resolution $h_i\to0$ with order $n$.}
\begin{ruledtabular} 
  \begin{tabular}{lllll}
  $h_i/m$&$m\eta=2.0$&$m\eta=1.0$&$m\eta=0.5$&$m\eta=0.0$\\%&order\\
\hline
1/100&0.54060	&0.54046	&0.54060	&0.54094	\\
1/120&0.54059	&0.54051	&0.54060	&0.54080	\\
1/140&0.54059	&0.54056	&0.54059	&0.54061	\\
$\to0$ &0.54059&0.54110&0.54066&0.53857\\
$n$ &1.58 &0.85 &2.24 &0.93\\
\hline
1/100&-0.19185&-0.19187&-0.19183&-0.19186\\%0&-0.86\\
1/120&-0.19184&-0.19185&-0.19184&-0.19185\\%&-0.15\\
1/140&-0.19183&-0.19184&-0.19184&-0.19184\\%&0.83\\
%$\to0$&-0.1918606&-0.1918089&0.0000000& \\
%order   &-2.5501262&1.9024536&0.0000000& \\
\end{tabular} \label{table:II}  
\end{ruledtabular}
\end{table}
%\end{widetext}

To evaluate the convergence rate with three resolutions $h_i$ with
$i=1,2,3$ we model the errors of a measured quantity $M_i=M(h_i)$
as $A_i.h^n_i$ in such a way that the extrapolated to infinite resolution
quantity $M_\infty=M(0)$ can be written as $M_\infty=M_i+\langle A\rangle h^n_i$, where
$\langle A\rangle$ is an averaged value of the $A_i$. Thus for the three resolutions
we have a system of three equations for the three unknowns,
$M_\infty$, $n$, and $\langle A\rangle$. $n$ representing the convergence rate and
$M_\infty$ the extrapolation to infinite resolution given in the tables
below.

The corresponding computation of radiated energy and angular momentum
from the waveforms extrapolated to an observer at infinity 
(from an extraction at $R=113m$) and summed
over all $(\ell,m)$-modes up to $\ell=6$ are
displayed in Tables \ref{table:III} and \ref{table:IV} showing consistent
approximate 3rd order convergence for the three resolutions n100, n120,
and n140. When using the extrapolated to infinite resolution horizon values as exact, the convergence order increases, 
and is over 4th order for the radiated angular momentum.  In all cases, the computations
are consistent in the first 3 digits. 
While taking as the exact reference the extrapolated to infinite resolution horizon values, the convergence 
is over 4th order. Consistent first 4-digits are computed in all cases.
Note that in all radiative computations we do not remove the initial 
data (spurious) radiation content to allow direct comparison with
the corresponding horizon quantities.

%\begin{widetext}
\begin{table}[h] 
  \caption{Energy radiated away in gravitational waves up to $\ell=6$
    for the $q=1/3$ nonspinning binary. Values for $m\eta=0.0$ are extrapolated. Convergence order calculated from the three resolutions, $n$, and using the two highest resolutions and assuming the converged value is the value calculated on the horizon, $n(AH)$.}
\begin{ruledtabular} 
  \begin{tabular}{llllll}
  $h_i/m$&$m\eta=2.0$&$m\eta=1.0$&$m\eta=0.5$&$m\eta=0.0$\\%&order\\
\hline
1/100&0.02017&0.02017&0.02017&0.02017\\%&0.26\\
1/120&0.02030&0.02030&0.02029&0.02030\\%&-0.81\\
1/140&0.02036&0.02036&0.02036&0.02036\\%&-5.62\\
$\to0$&0.02047&0.02045&0.02046& 0.02046\\
$n$   &3.03&3.23&3.08& 3.10\\
$n(AH)$&3.31&3.23&3.18& -----\\
\end{tabular} \label{table:III}  
\end{ruledtabular}
\end{table}
%\end{widetext}

In particular, very weak dependence on $\eta$ is found,
again as expected on the ground of gauge invariance of the
gravitational waveform extrapolated to an observer at infinite
location.

%\begin{widetext}
\begin{table}[h] 
  \caption{Angular momentum radiated away in gravitational waves up to $\ell=6$
    for the $q=1/3$ nonspinning binary. Values for $m\eta=0.0$ are extrapolated. Convergence order calculated from the three resolutions, $n$, and using the two highest resolutions and assuming the converged value is the value calculated on the horizon, $n(AH)$.}
\begin{ruledtabular} 
  \begin{tabular}{llllll}
$h_i/m$&$m\eta=2.0$&$m\eta=1.0$&$m\eta=0.5$&$m\eta=0.0$\\%&order\\
\hline
1/100&-0.19075&-0.19073&-0.19070&-0.19076\\%&-0.92\\
1/120&-0.19128&-0.19130&-0.19125&-0.19128\\%&-1.21\\
1/140&-0.19157&-0.19157&-0.19154&-0.19157\\%&-3.04\\
$\to0$&-0.19217&-0.19196&-0.19215& -0.19216\\
$n$   &2.56&3.38&2.62& 2.55\\
$n(AH)$&4.78&4.50&4.62& -----\\
\end{tabular} \label{table:IV}  
\end{ruledtabular}
\end{table}
%\end{widetext}

Tables \ref{table:III} and \ref{table:IV} also show that  
both radiative quantities, energy and angular momentum show very small
variations with respect to the
extrapolated $\eta\to0$ values, as expected from gauge invariant quantities.

Since the formula (\ref{eq:coordmom}) is not gauge invariant when applied to the
horizon of the final black hole we expect to find stronger variation
with $\eta$ when we use it to evaluate the linear momentum of the
remnant. We will pursue this exploration in more detail next in order
to assess what values of $\eta$ allow us to compute the recoil velocity
of the final black hole with a good accuracy. We are interested in particular,
for our typical numerical simulations resolutions, what values of $\eta$
can produce more accurate values of the recoil from the horizon by use of
(\ref{eq:coordmom}) than from the evaluation of the radiated linear
momentum at infinity.

\begin{figure}[h]
  \includegraphics[angle=0,width=1.0\columnwidth]{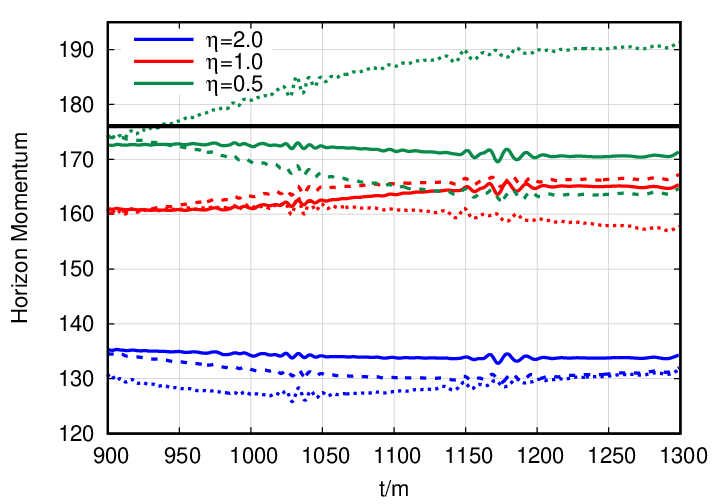}
  \caption{The horizon measure of the linear momentum (in km/s) after merger of a $q=1/3$ nonspinning binary for the three resolutions n100 (dotted), n120 (dashed), and n140 (solid) for $\eta=2/m$ (blue), $1/m$ (red), $0.5/m$ (green).
    The reference value of $V_f$ (black solid line) is found by extrapolation to infinite resolution of the radiated linear momentum.
\label{fig:etaconv}}
\end{figure}

Our starting point is the gauge choices that we have been using regularly
in our systematic studies of binary black hole mergers
($m\eta=2$ and Eqs.~(\ref{eq:gauge})) and numerical resolutions labeled
by the resolution at the extraction level of radiation as n100, n120, n140,
corresponding to wavezone resolutions of $h=1/1.00m, 1/1.20m, 1/1.40m$, respectively 
\cite{Healy:2017psd,Healy:2019jyf}. We use 10 levels of refinement with an outer boundary
at 400m.
For each of these three resolutions we add a set of simulations by decreasing
$\eta$ by factor of two, i.e. $\eta=1/m, 1/2m$.
The results of those
nine simulations are displayed in Fig.~\ref{fig:etaconv}. 
For $\eta=2/m$, the curves are very flat versus time after the
merger with the highest resolution run, n140, being notably so. However
their values for the evaluation of the recoil fall short compared to
the estimate coming from the extrapolation of the radiative linear
momentum to infinite resolution, represented by
the solid black lines at about 177km/s.

\begin{figure}[h]
  \includegraphics[angle=0,width=1.0\columnwidth]{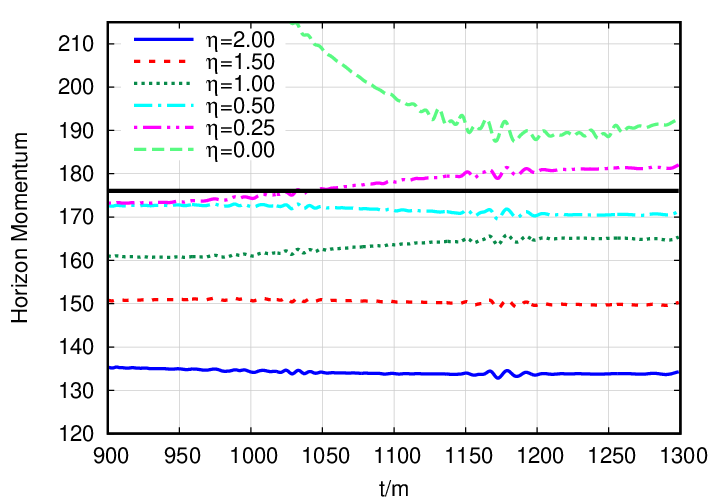}
  \caption{The horizon measure of the linear momentum (in km/s) after merger of a $q=1/3$ nonspinning binary lowering values of $\eta=2\to0$ at resolution n140. The reference value of $V_f$ (black solid line) is found by extrapolation to infinite resolution of the radiated linear momentum.
\label{fig:etaall}}
\end{figure}

The progression towards smaller $\eta$ shows closer agreement with that extrapolated value. The time dependence shows variations as we approach the smaller $\eta$ but still converging with resolution towards the expected 177km/s value and flatter for n140, but clearly the limit $\eta\to0$ requires much higher resolutions, as shown in Fig.~\ref{fig:etaall}. In our regime, reaching $\eta=1/m$ or $\eta=0.5/m$ seems a good compromise of accuracy versus cost of the simulation.

Fig.~\ref{fig:etaall} shows the progression of $m\eta=2.0, 1.5, 1.0, 0.5, 0.25$
for simulations with resolution n140. Notably they lie in a roughly linear
convergence towards the expected higher recoil velocity value 177km/s, but
as we reach the smaller $m\eta=0.25$ value it overshoots slightly, an effect
of the required higher resolution needed to resolve accurately smaller values
of $\eta$. In what follows we will restrict ourselves to values of
$m\eta=2.0, 1.0, 0.5$ to make sure we are in a convergence regime for
our standard resolutions n100, n120, n140. Note that we have verified that
the simulation with n140 and $\eta=0$ does not crash, but leads to inaccurate
results.

The radiation of linear momentum
in terms of the Weyl scalar $\psi_4$, as given by the formulas
in \cite{Campanelli:1998jv}, can be 
computed in a similar fashion as we compute
the energy and angular momentum radiated.  For this study,
we do not remove the initial burst of spurious radiation 
from the linear momentum calculation since we are interested
in comparing to the final velocity of the merged BH.  The burst
will impart a (usually) small kick to the center of mass of the system.
This allows direct comparison with horizon quantities in this paper.
For astrophysical applications the removal of the initial burst
of radiation is done in the waveform time domain and can be applied
to remove their contributions to the final mass, spin and recoil velocity.

Table \ref{table:V} shows that 
the radiation of linear momentum converges with resolution
(at an approximate 2.6-2.7th order) at similar
rates than the radiated energy and momentum (roughly 3rd order),
and still varies little with $\eta$. This is
expected on gauge invariance grounds.

%\begin{widetext}
\begin{table}[h] 
  \caption{Total linear momentum radiated in gravitational waves up to $\ell=6$
    for the $q=1/3$ nonspinning binary in km/s. Values for $m\eta=0.0$ are extrapolated. Convergence order $n$ is extrapolated to infinite resolution $h_i\to0$.
Bottom panel shows the angle (in degrees) of the net momentum with respect to the initial
x-axis.}
\begin{ruledtabular} 
  \begin{tabular}{lllll}
  $h_i/m$&$m\eta=2.0$&$m\eta=1.0$&$m\eta=0.5$&$m\eta=0.0$\\%&order\\
\hline
1/100&163.648&163.753&163.759&163.760\\%&3.99\\
1/120&168.569&168.678&168.660&168.662\\%&2.58\\
1/140&171.256&171.314&171.324&171.326\\%&2.58\\
$\to0$&176.750&176.422&176.708& 176.702\\
$n$   &2.582&2.699&2.608& 2.611\\
\hline
1/100&375.01\degree	&374.67\degree &373.61\degree	&-----	\\
1/120&374.93\degree 	&375.09\degree	&374.54\degree &-----	\\
1/140&375.50\degree	&375.59\degree &375.78\degree	&-----	\\
$\to0$&375.15\degree	&375.12\degree	&-----	&-----	\\
$n$   &1.99 	&2.00	&-----	&-----	\\
\end{tabular} \label{table:V}  
\end{ruledtabular}
\end{table}
%\end{widetext}

The values
extrapolated to infinite resolution lie in the 176-177km/s
range, consistently for all three values of $m\eta=2, 1, 0.5$.
Extrapolations of the recoil velocities to $\eta\to0$ are very close to
their values at $m\eta=2, 1, 0.5$ for
all three resolutions, again confirming the gauge independence of the results.
In addition we display the angle (in degrees, reversed sign) 
the net radiated momentum has with
respect to the x-axis (line joining the black hole initially).

%\begin{widetext}
\begin{table*}
  \caption{Horizon linear momentum measured at %$75m$ after merger
the interval $t=950m-1250m$ for the $q=1/3$ nonspinning binary in km/s.
The bottom panel gives the angle (in degrees) this magnitude subtends with the
initial x-axis. The measured standard deviation is given by the
$\pm$ in the relevant quantities. Convergence order and extrapolations are
given for $h_i\to0$ and $\eta\to0$.}
\begin{ruledtabular} 
  \begin{tabular}{llllll}
$h_i/m$&$m\eta=2.0$&$m\eta=1.0$&$m\eta=0.5$&$m\eta=0.0$&order\\
\hline
1/100&$137.515\pm1.12$&$160.507\pm1.14$&$171.977\pm3.99$&$183.397$&1.00\\
1/120&$139.543\pm0.86$&$161.615\pm1.36$&$174.703\pm2.80$&$193.765$&0.75\\
1/140&$139.194\pm0.40$&$165.014\pm1.55$&$174.948\pm0.87$&$181.161$&1.38\\
%$\to0$&0.0000000&160.1545809&174.9877581& & \\
%order   &0.0000000&-7.7959645&12.9339819& & \\
\hline
1/100&$373.83\degree\pm0.27\degree$& $370.26\degree\pm0.29\degree$& $367.52\degree\pm0.86\degree$&358.475\degree&0.38\\
1/120&$373.87\degree\pm0.48\degree$&$371.83\degree\pm0.27\degree$&$373.64\degree\pm0.36\degree$&372.909\degree	&25.41\\
1/140&$374.63\degree\pm0.58\degree$&$371.98\degree\pm0.09\degree$&$374.37\degree\pm0.16\degree$&373.389\degree&27.46\\
$\to0$&374.11\degree &372.36\degree	&375.81\degree	&373.418\degree	&\\
$n$   &2.00 	&6.07	&6.11	&18.53	&\\
  \end{tabular} \label{table:VI}  
\end{ruledtabular}
\end{table*}
%\end{widetext}

Table \ref{table:VI} displays the crucial result of recoil velocities
very close to their desired values for low resolutions when $m\eta=0.5$.
They do not vary so much with
resolution, as expected for horizon quantities,
when compared to the variations with respect
to the gauge choices. We observe close to a linear dependence on $\eta$ of
the recoil values. Their extrapolation to $\eta\to0$ overshoots the
expected value by a few percent, but the values at $m\eta=0.5$ are
nearly within 1\%. This provides an effective way to compute recoils,
since the corresponding radiative quantities are 3\% away for
n140. The horizon evaluations lying closer to the expected values by a
factor 3 over the radiative ones holds for all three
resolutions.
In addition we display the angle the recoil velocity subtends with
respect to the x-axis (line joining the black holes initially) showing
a notable agreement of this sensitive quantity with the results in
Table~\ref{table:V}.

The coordinate velocities do not benefit systematically from the
small $\eta$ gauges, but still provide a good bulk value as shown
in Table \ref{table:VII}. This shows the benefits of having a quasilocal
measure of the momentum of the hole over its horizon compared to the
local coordinate velocity of the puncture.

%\begin{widetext}
\begin{table}[h] 
  \caption{Coordinate trajectory velocity in km/s as
 measured at $575m$ after merger for the $q=1/3$ nonspinning binary.}
\begin{ruledtabular} 
  \begin{tabular}{lllll}
  $h_i/m$&$m\eta=2.0$&$m\eta=1.0$&$m\eta=0.5$&$m\eta=0.0$\\%&order\\
\hline
1/100&154.448&158.513&184.809&153.705\\%&-2.69\\
1/120&155.051&167.205&159.472&162.479\\%&0.65\\
1/140&158.123&165.702&165.957&165.966\\%&4.89\\
%$\to0$&154.3537630&0.0000000&0.0000000& \\
%order   &-10.9396618&0.0000000&0.0000000& \\
\end{tabular} \label{table:VII}  
\end{ruledtabular}
\end{table}
%\end{widetext}

\subsection{Validation for other mass ratios $(q=1/2,\ 1/5)$}\label{sec:validation}

In order to first validate our technique to extract the recoil velocity
of the remnant black hole from spinning binaries, we have considered
another unequal mass $(q=1/2)$
binary with initial separation $D=11m$. 
The resulting recoil will be along the orbital plane and due
entirely to the asymmetry produced by the unequal masses.
The results are presented in Table \ref{table:VIII}. 
Assuming the extrapolation to infinite resolution of the radiative
linear momentum computations is the most accurate one leads to
a recoil of $154.3\pm0.1$km/s. Even for the lowest computed resolution,
n100, the horizon evaluation for $m\eta=0.5$ at $159.6$km/s is a better
approximation
to that value than any radiative evolution at the same resolution
(145.9km/s), with errors of the order of $3\%$. This is also true
for the other two resolutions n120, and n140. Although,
as we have seen before, the improvement of those horizon
values are obtained by lowering the value of $m\eta\to0$,
rather than by higher resolution, as the horizon quasilocal measure has
essentially already converged at those resolutions.

\begin{table}[h] 
  \caption{Comparison of the computation of the recoil velocity (in km/s) of the
    remnant of a $q=1/2$, nonspinning binary by traditional radiation
    of linear momentum and the horizon formula (\ref{eq:coordmom})
    averaged between $t=1550m$ and $t=1850m$ for
    the traditional $\eta=2$ and for the $\eta=0.5$ case.
    Extrapolation to infinite resolution and convergence order is
  also given for the horizon and radiative extraction.
The bottom panel gives the angle (in degrees) of the recoil velocity
with respect to the x-axis.   
Standard deviations of horizon measurements are given as $\pm$ for
each quantity. Convergence with numerical resolution is also given.}
\begin{ruledtabular} 
  \begin{tabular}{lllll}
     & Radiation & Horizon & Radiation & Horizon \\
$h_i/m$&$m\eta=2.0$&$m\eta=2.0$&$m\eta=0.5$&$m\eta=0.5$\\
\hline
1/100 &    145.45 & $127.80\pm0.65$&  145.91&  $159.57\pm1.05$\\
1/120  &   149.45  &$121.02\pm0.79$ & 149.64 &$ 152.48\pm0.31$\\
1/140   &  151.38&  $118.77 \pm0.65$ &151.48  &$152.92\pm0.41$\\
$\to0$    & 154.28 & 117.08&  154.39& 150.98\\
$n$   &  3.31  &  5.49 &   3.18& 6.20\\
\hline
1/100 &384.43\degree  &$379.73\degree\pm1.35\degree$ &  390.60\degree   &$388.57\degree\pm0.55\degree$\\
1/120  &392.74\degree  &$387.84\degree\pm1.48\degree $& 391.04\degree &$ 388.75\degree\pm0.24\degree$\\   
1/140   &392.54\degree &$388.54\degree\pm1.41\degree$& 391.74\degree &$  389.80\degree\pm0.11\degree$\\   
$\to0$&394.75\degree &    390.48\degree & ---- &   -----\\ 
$n$   &6.17   &           6.14 & -----&-----\\
\end{tabular} \label{table:VIII}  
\end{ruledtabular}
\end{table}

Here we also provide the computation of the horizon
evaluations for the mass and spin of the remnant black hole 
in Table \ref{table:IX}. Those tables display the
excellent agreement  between the horizon and radiative
computation of the energy and angular momentum (with convergence
rates of the 3-4th order). It also displays the
agreement of those radiative computations for the $\eta=2$ and the
$\eta=0.5$ cases, as expected on the ground of gauge invariance at
the extrapolated infinite observer location. The table also shows the
robustness of the horizon computations at any of the used resolutions
(generally 5 digits) and an overconvergence due to those small differences.

\begin{table*}[t]
  \caption{Comparison of the computation of the horizon mass and spin of the
    remnant of a $q=1/2$, nonspinning binary with the radiation
    of the energy and angular momentum for the $\eta=2$ and for the
    $\eta=0.5$ cases.
    Extrapolation to infinite resolution and convergence order is
  also given for the horizon computation and the radiative extraction.}
\begin{ruledtabular} 
  \begin{tabular}{lllllll}
$m\eta=2.0$ & & & &\\
$h_i/m$ & $E_{rad}/m$ & $J_{rad}/m^2$ &  $m_f/m$  &  $\alpha_f/m^2$ & $\Delta m/m$ & $-\Delta J/m^2$\\
\hline
1/100& 0.02999& -0.30497& 0.96126& 0.62344 &0.03029&-0.30617\\
1/120 &0.03016 &-0.30600 &0.96125 &0.62345&0.03030&-0.30617\\
1/140& 0.03024& -0.30629& 0.96125& 0.62346&0.03030&-0.30617\\
$\to0$ &0.030337 &-0.30646 &0.96125 &0.62346&\\%0.0302684&-0.3061070\\
$n$  &   3.70        &    6.36        &    13.36   & 6.09&\\%0.86&0.98\\
\hline
$m\eta=0.5$ & & & &\\
%resolution & $E_{rad}$ & $J_{rad}$ &  $m_f$  &  $\alpha_f$& $\Delta m$ & $-\Delta J$\\
\hline
1/100& 0.02998& -0.30516& 0.96124& 0.62343&0.03031&-0.30620\\
1/120 &0.03014 &-0.30583 &0.96125 &0.62345&0.03030&-0.30617\\
1/140& 0.03022 &-0.30615& 0.96126& 0.62345&0.03030&-0.30617\\
$\to0$ &0.03033 &-0.30662 &0.96126&0.62345&\\%0.0302812&-0.3062583\\
$n$  &   3.42       &     3.39        &    7.20    & 8.62&\\%1.57&2.16\\
\end{tabular} \label{table:IX}  
\end{ruledtabular}
\end{table*}

%%%%%%%%
\begin{table*}[t]
  \caption{Comparison of the computation of the horizon mass and spin of the
    remnant of a $q=1/5$, nonspinning binary with the radiation
    of the energy and angular momentum for the $\eta=2$ and for the
    $\eta=0.5$ cases.
    Extrapolation to infinite resolution $h_i\to0$ and convergence order
$n$ is also given for the horizon computation and the radiative extraction.}
\begin{ruledtabular} 
  \begin{tabular}{lllllll}
$m\eta=2.0$ & & & &\\
$h_i/m$ & $E_{rad}/m$ & $J_{rad}/m^2$ &  $m_f/m$  &  $\alpha_f/m^2$& $\Delta m/m$ & $-\Delta J/m^2$\\
\hline
1/100 &0.01237& -0.15454 &0.98217 &0.41667&0.01253&-0.14804\\
1/120 &0.01225& -0.14872 &0.98235 &0.41667&0.01235&-0.14790\\
1/140 &0.01226& -0.14803 &0.98237 &0.41660&0.01232&-0.14796\\
$\to0$ &0.01227& -0.14788 &0.98238 &0.41667&\\%0.013458&-0.149908\\
$n$     &     11.59     &      11.41   &     10.78  & -15.49&\\%0.48&0.69\\
\hline
$m\eta=0.5$ & & & &\\
%resolution & $E_{rad}$ & $J_{rad}$ &  $M_f$  &  $\alpha_f$\\
\hline
1/100 &0.01310 &-0.16762 &0.98158& 0.41572&0.01320&-0.14944\\
1/120 &0.01241 &-0.15479 &0.98221& 0.41670&0.01249&-0.14800\\
1/140 &0.01227 &-0.14859 &0.98236& 0.41663&0.01234&-0.14795\\
$\to0$ &0.01221 &-0.13922 &0.98242& 0.41662&\\%-----&-----\\
$n$     &      8.12     &       3.30     &    7.59    &    14.39&\\%-----&-----\\
\end{tabular} \label{table:IXB}  
\end{ruledtabular}
\end{table*}

We complete our nonspinning studies by simulating a smaller mass ratio
$(q=1/5)$ binary with initial separation $D=10.75m$. 
The recoil from radiation of linear momentum extrapolates
to about 139km/s as shown in Table \ref{table:VIIIB}.
The horizon evaluation for $m\eta=2$ underevaluates this
by about 30\%, while for $m\eta=2$  the horizon formula is about 5\%
this value for the medium and high resolution runs. Given the smaller
mass ratio, the low resolution run is not as accurate.

We also provide as a reference the computation of the of the horizon
evaluations for the mass and spin of the remnant black hole 
in Table \ref{table:IXB}. Those tables display the
excellent agreement between the horizon and radiative
computation of the energy and angular momentum (with high convergence
orders). We find excellent
agreement of those radiative computations for the $\eta=2$ and the
$\eta=0.5$ cases, as expected from the gauge invariance of the waveforms
extrapolated to an infinite observer location. The table also shows the
robustness of the horizon computations at medium and high resolutions
(generally 3 digits) and an overconvergence due to those small differences.

\begin{table}[H] 
  \caption{Comparison of the computation of the recoil velocity (in km/s) of the
    remnant of a $q=1/5$, nonspinning binary by traditional radiation
    of linear momentum and the horizon formula (\ref{eq:coordmom})
    averaged between $t=2300m$ and $t=2500m$ for
    the traditional $\eta=2$ and for the $\eta=0.5$ case.
    Extrapolation to infinite resolution $h_i\to0$ and convergence order
$n$ is also given for the horizon and radiative extraction.
The bottom panel gives the angle (in degrees) of the recoil velocity
with respect to the x-axis.   
Standard deviations of horizon measurements are given as $\pm$ for
each quantity.}
\begin{ruledtabular}
  \begin{tabular}{lllll}
 & Radiation & Horizon & Radiation & Horizon \\
$h_i/m$&$m\eta=2.0$&$m\eta=2.0$&$m\eta=0.5$&$m\eta=0.5$\\
\hline
1/100  &   129.40 & $98.67\pm0.60$ &  130.21 & $209.00\pm2.16$\\
1/120  &   133.91 & $102.53\pm0.16$ & 133.19 & $131.93\pm1.22$\\
1/140  &   135.94 & $101.62\pm0.13$ & 136.70 & $143.28\pm0.17$\\
$\to0$  &   138.57 & 101.20 & -  &     146.78\\
$n$  &   3.72  &  7.46 &   -   &    10.16\\
\hline
1/100  &318.79\degree&  $317.09\degree\pm0.91\degree$&333.20\degree& $366.32\degree\pm0.56\degree$ \\  
1/120  &374.99\degree&  $376.95\degree\pm0.33\degree$&345.29\degree&  $  293.68\degree\pm0.04\degree$ \\  
1/140  &332.13\degree&  $329.04\degree\pm0.33\degree$&385.40\degree&  $  384.90\degree\pm0.11\degree$ \\  
%$\to0$& \\
%$n$  & \\
\end{tabular} \label{table:VIIIB}  
\end{ruledtabular}
\end{table}

\begin{table*}[t] 
  \caption{Comparison of the computation of the recoil velocity (in km/s) of the
    remnant of a $q=1$, $\alpha_i=\pm0.8$ binary by traditional radiation
    of linear momentum and the horizon formula (\ref{eq:coordmom})
    averaged between $t=1050m$ and $t=1350m$ for
    the traditional $\eta=2$ and the $\eta=1\text{ and }0.5$ cases.
    Extrapolation to infinite resolution $h_i\to0$ and convergence order
$n$ is also given for the radiative extraction.
The bottom panel gives the angle (in degrees) of the recoil velocity
with respect to the x-axis.   
Standard deviations of horizon measurements are given as $\pm$ for
each quantity. In this case $m\eta=1$ produces a better measure of recoil velocity
        than other choices of $m\eta$.}
\begin{ruledtabular} 
  \begin{tabular}{lllllll}
 & Radiation & Horizon& Radiation&Horizon & Radiation & Horizon \\
$h_i/m$&$m\eta=2.0$&$m\eta=2.0$&$m\eta=1.0$&$m\eta=1.0$&$m\eta=0.5$&$m\eta=0.5$\\
\hline
1/100  &   387.92 & $331.98\pm1.50$ &-&-& 388.13 & $419.89\pm0.30$\\
1/120  &   394.16 & $331.84\pm1.41 $&394.13&396.41$\pm$0.22& 394.16 & $421.02\pm0.38$\\
1/140  &   397.43 & $331.75\pm1.49 $&-&-& 397.16 & $419.67\pm0.86$\\
$\to0$ & 403.42 & 331.40 &-&-& 402.00 &  - \\
$n$    & 2.82   &  1.46  &-&-&  3.12  &  - \\
\hline
1/100  &135.51\degree&$135.81\degree\pm0.93\degree$&-&-& 135.70\degree&  $137.25\degree\pm0.27\degree$ \\  
1/120  &136.70\degree& $ 135.72\degree\pm0.92\degree$&136.65\degree&135.51\degree$\pm$0.27\degree& 136.72\degree& $ 136.65\degree\pm0.14\degree $\\  
1/140  &137.39\degree&$  135.80\degree\pm0.94\degree$&-&-& 137.15\degree& $  136.61\degree\pm0.13\degree $\\  
$\to0$&139.07\degree& 135.78\degree&- &-& 137.60\degree&  136.83\degree \\
$n$  &2.23&              4.26&  - &-&            4.23&              4.29 \\
\end{tabular} \label{table:X}  
\end{ruledtabular}
\end{table*}

\begin{table*}[t]
  \caption{Comparison of the computation of the horizon mass and spin of the
    remnant of a $q=1$, $\alpha_i=\pm0.8$ binary with the radiation
    of the energy and angular momentum for the $\eta=2$ and for the
    $\eta=0.5$ cases.
    Extrapolation to infinite resolution
$h_i\to0$ and convergence order $n$ is
  also given for the radiative extraction.}
\begin{ruledtabular} 
  \begin{tabular}{lllllll}
$m\eta=2.0$ & & & &\\
\hline
$h_i/m$ & $E_{rad}/m$ & $J_{rad}/m^2$ &  $m_f/m$  &  $\alpha_f/m^2$& $\Delta m/m$ & $-\Delta J/m$\\
1/100 & 0.03872 &-0.34232 &0.95071 &0.68413&0.03936&-0.34403\\
1/120 & 0.03898 &-0.34320 &0.95071 &0.68413&0.03936&-0.34403\\
1/140 & 0.03911 &-0.34362 &0.95071 &0.68413&0.03936&-0.34404\\
$\to0$ & 0.03930 &-0.34421 &0.95071&0.68413&-&-\\%0.0393697&-0.3440342\\
$n$ &    3.28         & 3.45  & - & -&-&-\\%1.39&2.00\\%6.62    & -14.73\\
\hline
$m\eta=1.0$ & & & &\\
\hline
1/120 & 0.038979 &-0.34318 &0.95071 &0.68413 & 0.03936 & 0.34404 \\
\hline
$m\eta=0.5$ & & & &\\
\hline
1/100 & 0.03872 &-0.34232 &0.95071 &0.68413&0.03935&-0.34405\\
1/120 & 0.03898 &-0.34318 &0.95071 &0.68413&0.03936&-0.34403\\
1/140 & 0.03911 &-0.34358 &0.95071 &0.68413&0.03935&-0.34404\\
$\to0$ & 0.03930 &-0.34411 &0.95071 &0.68413&-&-\\%0.0392901&-0.3443170\\
$n$ &     3.30     &       3.58      & - &-&-&-\\%2.16&0.91\\%10.98     &      12.93\\
\end{tabular} \label{table:XI}  
\end{ruledtabular}
\end{table*}

\subsection{Spinning black holes}\label{sec:spin}

In order to further verify our technique to extract the recoil velocity
of the remnant black hole from spinning binaries, we have considered
an equal mass $(q=1)$ binary with spins $(\alpha_{1,2}=\pm0.8)$
antialigned with the orbital angular momentum. This system has
an initial separation of $D=10m$. 
The resulting recoil will be along the orbital plane and due
entirely to the asymmetry produced by the opposing spins.
The results are presented in Table \ref{table:X}, showing that
assuming the extrapolation to infinite resolution of the radiative
linear momentum computations is the most accurate one, leading to
a recoil of $403\pm1$km/s, even for the lowest computed resolution,
n100, the horizon evaluation for $m\eta=0.5$ at $420$km/s is as good
to that value than any radiative evolution at the same resolution
(388km/s), with errors of the order of $4\%$.
As we have seen before, the improvement of those horizon
values are obtained by lowering the value of $m\eta\to0$,
rather than by higher resolution, as the horizon quasilocal measure has
essentially already converged at those resolutions.

For the sake of completeness, and to verify the accuracy of the horizon
evaluations for the mass and spin of the remnant black hole, we provide
their computation in Tables \ref{table:XI}. Those tables display the
excellent agreement  between the horizon and radiative
computation of the energy and angular momentum (with convergence
rates of the 3-4th order). It also displays the
agreement of those radiative computations for the $\eta=2$ and the
$\eta=0.5$ cases, as expected on the ground of gauge invariance at
the extrapolated infinite observer location. Further, the tables show the
robustness of the horizon computations at any of the used resolutions
(generally 5 digits).

A final note on the convergence studies carried out in this section is
that we observe a good convergence rate of 3rd to 4th order for radiative
quantities while for horizon quantities a more wider range of values,
with sometimes overconvergence. This is due to the fact that the horizon
evaluations (particularly for the mass and spin) lead to very accurate
values and hence small differences between the three resolutions chosen
for the simulations (n100, n120, and n140). In order to seek very significant
differences between resolutions, factors larger than 1.2 should be chosen
(although requiring much larger computational resources). Note that
nevertheless we have been able to prove that horizon quantities
can be evaluated very accurately at any of the resolutions quoted above.

\section{Other Gauges Studies}\label{sec:OtherStudies}
\begin{table*}[t]
  \caption{Comparison of the computation of the recoil velocity (in km/s) of the
    remnant of a $q=1/3$, $\alpha_i=\pm0$ binary by traditional radiation
    of linear momentum and the horizon formula (\ref{eq:coordmom})
    measured $75m$ after merger. Computation uses
    the $\partial_0$-gauge $\eta=2$ case for resolutions n100
	n120 and n140 and $\partial_0$-gauge $\eta=1,0.5$ cases for resolution n140.
    Extrapolation to infinite resolution $h_i\to0$ and 
convergence order $n$ is
  also given for the radiative extraction for appropriate cases.
The bottom panel gives the angle (in degrees) of the recoil velocity
with respect to the x-axis.   
Standard deviations of horizon measurements are given as $\pm$ for
each quantity. Compare this results with those of Tables~\ref{table:V}-\ref{table:VI}.   }
\begin{ruledtabular} 
  \begin{tabular}{lllllll}
 & Radiation & Horizon & Radiation & Horizon & Radiation & Horizon \\
$h_i/m$&$m\eta=2.0$&$m\eta=2.0$&$m\eta=1.0$&$m\eta=1.0$&$m\eta=0.5$&$m\eta=0.5$\\
\hline
1/100  &   163.67 & $127.11\pm0.87$ &-----  &----- &-----&-----\\
1/120  &   168.57 & $129.10\pm0.98 $&-----  &----- &-----&-----\\
1/140  &   171.24 & $128.84\pm0.57 $& 171.30 & $152.11\pm1.92$&171.33&$160.9\pm1.55$\\
$\to0$ & 176.65 & 129.43&-----  &-----   &-----&-----\\
$n$    & 2.60   &  6.12 &-----   &----- &-----&----- \\
\hline
1/100  &375.82\degree&$369.58\degree\pm0.29\degree$&----- &-----  &-----&----- \\  
1/120  &375.17\degree& $ 368.99\degree\pm0.53\degree$&----- &----- &-----&-----\\  
1/140  &375.63\degree&$  369.14\degree\pm0.63\degree$&375.67\degree& $ 371.55\degree\pm0.09\degree $&375.88\degree&$373.93\degree\pm0.10\degree$\\  
$\to0$&-----&369.24\degree&----- &----- &-----&-----\\
$n$  &-----&4.14  &-----&-----&-----&-----\\
\end{tabular} \label{table:XII}  
\end{ruledtabular}
\end{table*}

Right after the breakthrough that allowed evolving binary black holes
with the moving puncture formalism \cite{Campanelli:2005dd,Baker:2005vv},
several papers analyzed extensions of the basic gauges (\ref{eq:gauge}) -
(\ref{eq:gaugenasa}).
In Refs. \cite{Gundlach:2006tw} and \cite{vanMeter:2006vi} several parametrizations of the shift
conditions are studied, and displayed some (slight) preference for the
$\partial_0$-gauge over the $\partial_t$-gauge
(See Table I of Ref.~\cite{Gundlach:2006tw} and Fig. 10 of Ref.~\cite{vanMeter:2006vi}).

The sensitivity of the computed recoil on the gauge give us an opportunity
to quantify the relative accuracy of the $\partial_0$-gauge
versus the $\partial_t$-gauge.

We will also exploit the possibility of using a variable $\eta(x^k(t))$ to obtain
both, the benefits of accuracy around the black holes and a good coordinate
behavior connecting the horizon results with asymptotia.

\subsection{$\partial_0$-gauge}\label{sec:d0}
\begin{table*} 
  \caption{Top panel shows the comparison of the computation of the recoil velocity (in km/s) of the
    remnant of a $q=1$, $\alpha_i=\pm0.80$ binary by traditional radiation
    of linear momentum and the horizon formula (\ref{eq:coordmom})
    averaged from $t/m = 1050$ to 1350 using 
    the $\partial_0$-gauge with $\eta=2$ and $\eta=0.5$ case at resolutions n100 and
	n120.
The bottom panel gives the angle (in degrees) of the recoil velocity
with respect to the x-axis. Both panels give also the $\partial_t$ 
results from Table \ref{table:X} for comparison.
Standard deviations of horizon measurements are given as $\pm$ for
each quantity.}
\begin{ruledtabular} 
  \begin{tabular}{llllll}%llll}
 &Gauge & Radiation & Horizon & Radiation & Horizon \\
$h_i/m$&$\partial_0,\partial_t$&$m\eta=2.0$&$m\eta=2.0$&$m\eta=0.5$&$m\eta=0.5$\\
\hline
1/100&$\partial_0$  & 387.94 & $294.93\pm2.43$&388.01&$377.37\pm1.56$\\
1/100&$\partial_t$  &   387.92 & $331.98\pm1.50 $& 388.13 & $419.89\pm0.30$\\
1/120&$\partial_0$  & 394.18 & $294.82\pm2.41$&394.13&$376.89\pm0.94$\\
1/120&$\partial_t$  &   394.16 & $331.84\pm1.41 $& 394.16 & $421.02\pm0.38$\\
\hline
1/100&$\partial_0$  &135.53\degree& $ 135.71\degree\pm1.00\degree $&135.78\degree&$137.31\degree\pm0.27\degree$\\  
1/100&$\partial_t$  &135.51\degree& $ 135.81\degree\pm0.93\degree$& 135.70\degree& $ 137.25\degree\pm0.27\degree $\\
1/120&$\partial_0$  &136.71\degree& $ 135.62\degree\pm1.00\degree $&136.69\degree&$136.65\degree\pm0.18\degree$\\  
1/120&$\partial_t$  &136.70\degree& $ 135.72\degree\pm0.92\degree$& 136.72\degree& $ 136.65\degree\pm0.14\degree $\\
\end{tabular} \label{table:XIIB}  
\end{ruledtabular}
\end{table*}

\begin{figure}[h]
  \includegraphics[angle=0,width=1.0\columnwidth]{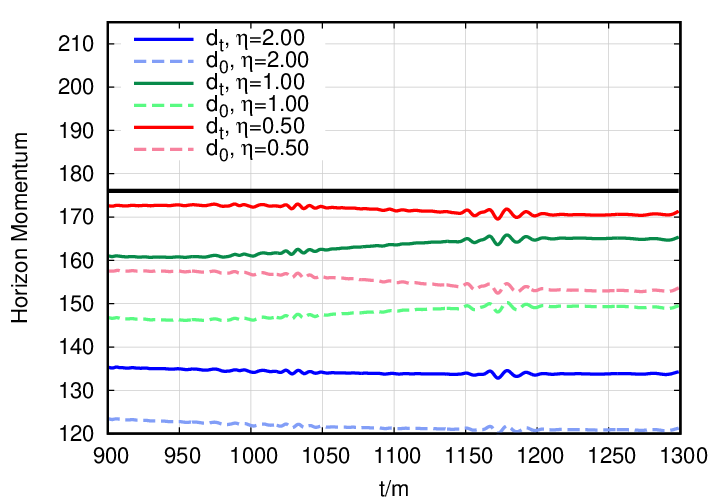}
  \caption{Comparative results of the $\partial_t$-gauge (solid) and $\partial_0$-gauge (dashed)
for the horizon measure of the linear momentum (in km/s) after merger of a $q=1/3$ nonspinning binary for the n140 resolution for $\eta=2/m$ (blue), $1/m$ (green), $0.5/m$ (red).
    The reference value of $V_f$ is found by extrapolation to infinite resolution of the radiated linear momentum.
\label{fig:etad0dt}}
\end{figure}

In Fig. \ref{fig:etad0dt} we draw a comparative analysis of
the final black hole horizon recoil as computed in the
$\partial_0$ and $\partial_t$ gauges, at our highest resolution n140.
In all three cases, $\eta=2,1,0.5$, the computation in the
$\partial_t$-gauge is notably and systematically closer to the
expected ($V_f\sim177$km/s) recoil velocity. This also provides a
scale of the accuracy of the evaluation of the recoil for
our new preferred value, $\eta=0.5$.

\begin{figure}[h]
  \includegraphics[angle=0,width=1.0\columnwidth]{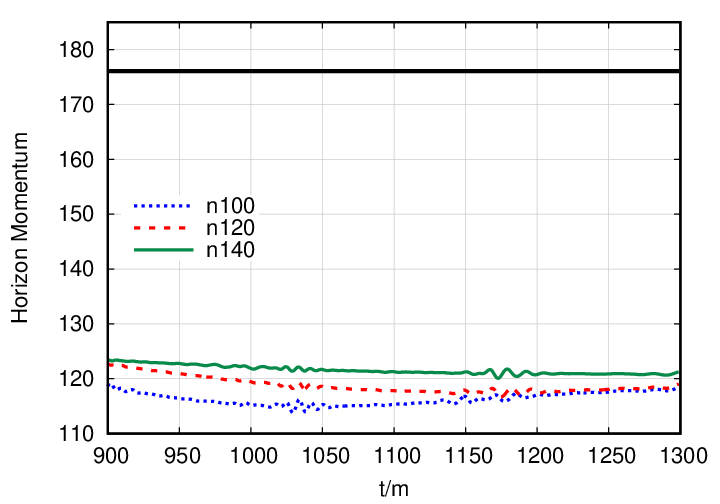}
  \caption{The horizon measure of the linear momentum (in km/s) after merger of a $q=1/3$ nonspinning binary for the three resolutions labeled as n100, n120, and n140 in the $\partial_0$-gauge for $\eta=2/m$.
    The reference value of $V_f$ is found by extrapolation to infinite resolution of the radiated linear momentum in this $\partial_0$-gauge.
\label{fig:d0etaconv}}
\end{figure}

Convergence with resolution does not resolve this discrepancies in
favor of the $\partial_t$-gauge as displayed in
Fig.~\ref{fig:d0etaconv} for $m\eta=2$ in
in the $\partial_0$-gauge at n100, n120, and n140 resolutions.
With the limit $\eta\to0$ being even harder to resolve than in
the $\partial_t$-gauge case.

%\begin{figure}
%  \includegraphics[angle=0,width=\columnwidth]{figs/D9_Etad0All_hn_mom3_vs_time.png}
%  \caption{The horizon measure of the linear momentum (in km/s) after merger of a $q=1/3$ nonspinning binary lowering values of $m\eta=2\to0$ at resolution n140 for the $\partial_0$-gauge. The reference value of $V_f$ is found by extrapolation to infinite resolution of the radiated linear momentum.
%\label{fig:d0etaall}}
%\end{figure}

Those results for the evolutions in the $\partial_0$-gauge
are summarized in Table~\ref{table:XII}
were we should compare its results with those in Tables~\ref{table:IV}-\ref{table:V}
in the $\partial_t$-gauge. While the computation of the extracted radiation are
comparable and convergent to essentially the same values, i.e. a recoil
magnitude of about 177km/s and an angle with the $x$-axis of $375\degree$,
the closeness to those values in the $\partial_t$-gauge is apparent
for all values of $\eta$.
A second control case is studied in Table~\ref{table:XIIB},
were we consider the spinning binary system described in Section~\ref{sec:spin}.
We directly compare the $\partial_0$-gauge new simulations with the
$\partial_t$-gauge simulations reported in Table~\ref{table:X}.
Based on those results we expect recoil magnitudes in the 402-403km/s
and angles in the $138\degree-139\degree$ ranges.
The results for $\eta=2$ confirms the closer to the expected recoils
in the $\partial_t$-gauge, and those of $\eta=0.5$ bracket it, with
preference for the $\partial_t$-gauge. We also observe again that
the horizon values are much more sensitive to the values of $\eta$
than to resolutions (at least for these 1.2 increase factors).

In conclusion we observe the advantage of working in the $\partial_t$-gauge 
over the $\partial_0$-gauge regarding computation of recoil velocities from
the horizon formula~(\ref{eq:coordmom}). This agrees with our generic experience for
binary black holes simulations being more accurate in our standard
$\partial_t$-gauge at the same resolutions and same values of $\eta$,
but now we have quantified it in the recoil computations example.
As we converge to higher resolutions, both gauges lead to consistent 
and accurate solution in all studied quantities.

\subsection{$\eta$-variable gauge}\label{sec:etar}
\begin{table*} 
  \caption{Comparison of the computation of the recoil velocity (in km/s) of the
    remnant of a $q=1/3$, $\alpha_i=0$ binary by traditional radiation
    of linear momentum and the horizon formula (\ref{eq:coordmom})
    measured $75m$ after merger for
    the modified N12 and N10 gauges. The standard $\eta=1$ from 
Tables~\ref{table:V} and \ref{table:VI} is also provided for reference.
    Extrapolation to infinite resolution $h_i\to0$ and
convergence order $n$ is
  also given for the radiative extraction.
The bottom panel gives the angle (in degrees) of the recoil velocity
with respect to the x-axis.   
Standard deviations of horizon measurements are given as $\pm$ for
each quantity.   }
\begin{ruledtabular} 
  \begin{tabular}{lllllll}
     & Radiation & Horizon & Radiation & Horizon &Radiation&Horizon\\
$h_i/m$&N12&N12&N10&N10&$m\eta=1$&$m\eta=1$\\
\hline
1/100  &   163.65 & $169.88\pm1.29$ & 163.71 & $150.54\pm1.18$&163.75&160.51$\pm$1.14\\
1/120  &   168.59 & $172.56\pm0.75 $& 168.61 & $153.52\pm0.86$&168.66&161.62$\pm$1.36\\
1/140  &   171.20 & $177.64\pm1.78 $& 171.26 & $155.31\pm1.75$&171.31&165.01$\pm$1.55\\
$\to0$ & 176.09 & ----- & 176.52 &  160.14 &176.42 &-----\\
$n$    & 2.78   &  ----- &  2.64  &  2.04 &2.70 &-----\\
\hline
1/100  &373.01\degree&$371.21\degree\pm0.93\degree$& 374.39\degree&  $364.97\degree\pm0.27\degree$ &374.67\degree&370.26\degree$\pm$0.29\degree\\  
1/120  &374.02\degree& $ 373.17\degree\pm0.92\degree$& 374.85\degree& $ 366.81\degree\pm0.14\degree $&375.09\degree&371.83\degree$\pm$0.27\degree\\  
1/140  &374.72\degree&$  377.24\degree\pm0.94\degree$& 375.57\degree& $  367.90\degree\pm0.13\degree $&375.59\degree&371.98\degree$\pm$0.09\degree\\  
$\to0$&378.40\degree& 377.24\degree&  -----&  -----&375.12\degree&372.36\degree\\
$n$  &1.13&              2.15&  -----     &   -----&2.00&6.07\\
\end{tabular} \label{table:XIII}  
\end{ruledtabular}
\end{table*}

%\subsection{$\eta$-variable gauge}\label{sec:etar}
\begin{table*}[t] 
  \caption{Comparison of the computation of the recoil velocity (in km/s) of the
    remnant of a $q=1$, $\alpha_i=\pm 0.80$ binary by traditional radiation
    of linear momentum and the horizon formula (\ref{eq:coordmom})
    measured from $t/m=1050\text{ to } 1350$ for
    the modified N12 and N10 gauges using $\eta\to1$ asymptotically, as well as $m\eta=1.0$ for comparison.
The bottom panel gives the angle (in degrees) of the recoil velocity
with respect to the x-axis.   
Standard deviations of horizon measurements are given as $\pm$ for
each horizon quantity.   }
\begin{ruledtabular} 
  \begin{tabular}{lllllll}
     & Radiation & Horizon & Radiation & Horizon & Radiation & Horizon \\
$h_i/m$&N12&N12&N10&N10&$m\eta=1.0$&$m\eta=1.0$\\
\hline
1/120  &   394.88 & $379.66\pm0.80 $& 393.87 & $335.74\pm1.10$&394.13&396.41$\pm$0.22\\
\hline
1/120  &137.44\degree& $ 141.54\degree\pm0.46\degree$& 136.38\degree& $ 128.90\degree\pm0.08\degree $&136.65\degree&135.51\degree$\pm$0.27\degree\\  
\end{tabular} \label{table:XIIIB}  
\end{ruledtabular}
\end{table*}

In addition to the changes in the (constant) values of $\eta$,
different functional dependences for $\eta(x^k,t)$ have been
proposed in 
\cite{Zlochower:2005bj,Mueller:2009jx, Mueller:2010bu, Schnetter:2010cz,Alic:2010wu,Lousto:2010ut}. %, with diverse motivations.

Here we use a modified form motivated by the results of \cite{Krishnan:2007pu}
and this paper that the recoil velocities (of a merged binary)
are more accurately 
computed when using the quasilocal horizon measure of the momentum
with smaller $\eta$ and that the generic evolution is more accurate and
convergent for larger values of $\eta$.

We hence propose a simple variant for comparable masses binary
\beq\label{eq:eta}
m\eta(x^k,t)=m\eta_\infty-A\,e^{-r^2/s^2}
\eeq
where $m=m_1+m_2$ and $r=|\vec{r}-\vec{r}_{com}|$ 
is the distance from the (Newtonian) center of mass of the system
[PN corrections could be added if needed \cite{Bernard:2017ktp}],
\beq
\vec{r}_{com}=(m_1\,\vec{x}_1+m_2\,\vec{x}_2)/m
\eeq
where $\vec{x}_1(t)$ and  $\vec{x}_2(t)$ are the punctures location,
and $s$ is a width of the Gaussian that can be conveniently chosen, for instance, $s=2m$. Typically for our simulations normalization is chosen such
that $m\sim1$ (and $M_{ADM}<1$).

The choice to center the Gaussian correction to the $\eta_\infty$ at the
center of mass is to provide a simple way of following the final
black hole after merger, even if acquired a large recoil velocity.
The motivation to set different values around the black hole is
to provide enough accuracy and convergence in the strong field regime,
while preserving the benefits of the coordinates adapted to recoil
measurements away from the remnant hole.

In order to assess those statements we have considered two cases labeled
as N10 and N12, respectively determined by the $-$ and $+$ in 
\beq\label{eq:Neta}
m\eta(x^k,t)=1\mp\,e^{-r^2/(2m)^2}
\eeq
with the same reference asymptotic value of 1 at infinity and vanishing or
taking the standard value of 2 at the (final) hole location.

The simulations (in the standard $\partial_t$-gauge)
make use of this (\ref{eq:Neta}) dependence during the
whole run, not only during the post-merger phase, but the horizon of
the final black hole is only found and evaluated for linear momentum
after the merger occurs (about $t\sim725m$).

\begin{figure}[h]
  \includegraphics[angle=0,width=1.0\columnwidth]{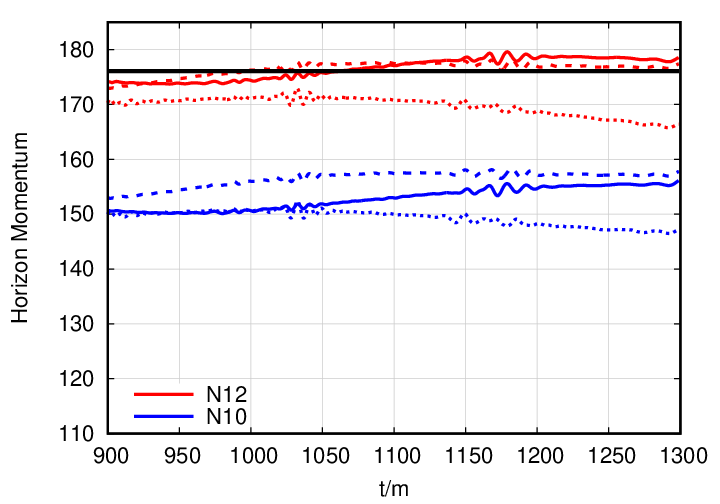}
  \caption{The horizon measure of the linear momentum after merger of a $q=1/3$ nonspinning binary for the three resolutions labeled as n100 (dotted), n120 (dashed), and n140 (solid) for $\eta=$ N12 (red) and N10 (blue), top to bottom respectively.
The reference value of $V_f$ is found by extrapolation to infinite resolution of the radiated linear momentum.
\label{fig:etaN12N10conv}}
\end{figure}

The results of the two cases are displayed in Fig. \ref{fig:etaN12N10conv}.
Each case has been studied at our standard n100, n120, n140 resolutions to
convey an idea of the convergence. The upper panel displays a very good agreement with the expected recoil, particularly for the medium and high resolutions.
The agreement is even better than the constant $m\eta=1$ case, displayed
in the central panel of Fig.~\ref{fig:etaconv}, which lies in between the
N12 and N10 cases.
To confirm the improvements reached by this variable-$\eta$ gauge, the
case N10, where the asymptotic value is the same, equal to 1, but where
the $\eta$ is reduced near the black hole, reduces notably the accuracy necessary
to compute the linear momentum of the horizon and convergence is still more
challenging than in the previous N12 case.

Those results for the variable $\eta$-gauge, as in (\ref{eq:Neta})
are summarized in Table~\ref{table:XIII}
were we compare its results with those in Tables~\ref{table:IV}-\ref{table:V}
in the $\eta=1$-gauge. While the computation of the extracted radiation are
comparable and convergent to essentially the same values, i.e. a recoil
magnitude of about 176-7km/s and an angle with the $x$-axis of 375-6\degree,
the closeness to those values in the N12-gauge is apparent
followed by the $m\eta=1$ (the reference value) and lagged by the N10-gauge,
indicating that while the same asymptotic $m\eta=1$ value is shared by
the three gauges, that with $m\eta\to2$ near the horizon of the black hole
produces the most accurate results for the recoil computed via the
horizon formula (\ref{eq:coordmom}).

A second control case is studied in Table~\ref{table:XIIIB},
were we consider the spinning binary system described in Section~\ref{sec:spin}.
We directly compare the N12-N10-gauge new simulations with each other 
using the radiation values as the more accurate references, and we find
again the confirmation that the N12 results are much closer to the
expected results than the N10 ones.

In conclusion, a first exploration of an $\eta$-variable leads to immediate
benefits and opens the possibilities for further refinement of its parameters
to have both, accuracy and precision improved in the numerical simulations 
of merging binary black holes.

\subsection{Small mass ratio}\label{sec:smallq}

Because of the technical similarities (although with complementary mass ratio applications), here we briefly discuss dealing with the small mass ratio binaries by the modeling of the damping parameter $\eta$.
The proposal for $\eta$ in Eq. (\ref{eq:eta}) is meant to be used for comparable masses $q>1/10$ when we can still use a constant $\eta$ for evolutions. For smaller mass ratios this $\eta_\infty$ can be replaced by the $\eta(W)$
(the conformal factor $W=\sqrt{\chi}=\exp(-2\phi)$ suggested by~\cite{Marronetti:2007wz})
used in Ref. \cite{Lousto:2010ut} or a modification of it given below or yet other based on superposition of weighted Gaussians.
Note that the recoil for $q<1/10$ is small.
This question has been already studied in \cite{Mueller:2009jx,Lousto:2010tb,Muller:2010zze,Alic:2010wu,Schnetter:2010cz}

Here we simply bring back some of those ideas, assuming we evaluate $\eta(\vec{r}_1(t),\vec{r}_2(t))$ parametrized by the black holes 1 and 2 punctures trajectories $(\vec{r}_1(t),\vec{r}_2(t))$

The (initial) conformal factor evaluated at every time step is
\beq
\psi_0=1+{\frac {m_1}{{|\vec{r}-\vec{r}_1(t)|}}}+{\frac {m_2}{{|\vec{r}-\vec{r}_2(t)|}}}
\eeq

And we can then define analogously to the $\eta(W)$
\beq\label{eq:eta0}
m\eta_\psi=A+ B\,{\frac {\sqrt {|\vec{\nabla}_r\psi_0|^2}}{ \left( 1-{\psi_0}^{a} \right) ^{b}}} .
\eeq

\begin{figure}[h]
\includegraphics[angle=0,width=0.7\columnwidth]{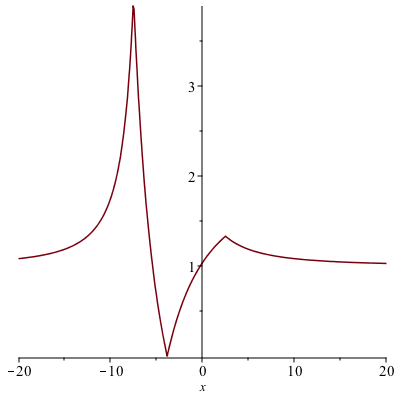}
\caption{$\eta_\psi$ profile for ($m=m_1+m_2=1$ here)
$m_1=1/4$, $m_2=3/4$; $x_1= -7.5$, $x_2=2.5$; $a=1$, $b=2$; $A=1$, $B=1$.
\label{fig:eta0}}
\end{figure}

We plot an example in Fig.~\ref{fig:eta0}. At the puncture $m\eta=2$ and at the center of mass $m\eta=2.04$, but it goes through a minimum $m\eta=0$ and a this point, as well as the punctures, $\eta$ is $C^0$. Since the gauge condition Eq.~(\ref{eq:gauge}) involves an integration, this might still be fine
for evolutions (as well as at the punctures).

\begin{figure}[h]
\includegraphics[angle=0,width=0.7\columnwidth]{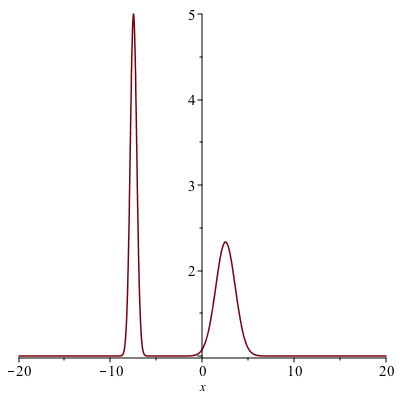}
\caption{$\eta_G$ profile for ($m=m_1+m_2=1$ here)
  $m_1=1/4$, $m_2=3/4$; $x_1= -7.5$, $x_2=2.5$; $A=1$, $B=1$, $C=1$;
  $s_1=2\,m_1$, $s_2=2\,m_2$
\label{fig:etaG}}
\end{figure}

A second alternative smoother behavior is the superposition of Gaussian (See also \cite{Muller:2010zze})
\beq\label{eq:etaG}
\eta_G=\frac{A}{m}+{\frac {B}{m_1}{e^{-|\vec{r}-\vec{r}_1(t)|^{2}/{s_1}^{2}}}}+{\frac {C}{m_2}{e^{-|\vec{r}-\vec{r}_2(t)|^{2}/s_2^{2}}}},
\eeq
which for the parameters of the previous example we display in Fig. \ref{fig:etaG}, behaving like 1.25 at the first puncture, 1.75 at the second and is essentially 1 in between and far away from the binary.

The explicit application and evaluations in actual simulations of small
mass ratio binary black hole mergers is left for an independent study
(Healy et al. 2020b, paper in preparation).

%%%%%%%
\section{Discussion}\label{sec:dis}

The purpose of this study was to assess what choices of $\eta$ lead to
accurate measures of the linear momentum of the horizon with the
non-gauge-independent formula (\ref{eq:coordmom}), and we found that for small
values $m\eta\leq0.5$ this is a reliable measure and can compete with the
measurement at $\scri^+$ of the radiated momentum carried by the gravitational
waves. As with the computations of the mass and angular momentum of
the remnant via the horizon measure and at infinity, it is important to
have two concurrent methods to assess errors of those measures.
Further accuracy could be achieved by the use of a variable $\eta$
(See Eq. (\ref{eq:eta}). Our results indicate that the choice of
$m\eta=2$ at the horizon, with lower values at asymptotically far distances
from the source(s), produce
the best results for evaluation of the recoil.

While these gauges were studied in detail for a nonspinning $q=1/3$ binary and
verified as control cases
for the $q=1/2$ and $q=1/5$ binaries as well as for a $q=1$ spinning case,
we expect these conclusions
to be general and plan to apply the findings to simulations where the
computation of recoil is important, including precessing binaries.
The cross checking with radiated linear momentum will provide a control
to its applicability and can be carried out concurrently in each simulation.

Finally, we have been able to assess the relative accuracy of the two
original moving puncture choices for the shift, the $\partial_t$ and
$\partial_0$ gauges regarding their accuracy to evaluate the recoil
and found that the $\partial_t$-gauge seems to be superior, at the
current typical numerical resolutions.

%%%%%%%

\begin{acknowledgments}
The authors thank Y.Zlochower for discussions on the gauge choices.
The authors gratefully acknowledge the National Science Foundation (NSF)
for financial support from Grants
No.\ PHY-1912632, No.\ PHY-1707946, No.\ ACI-1550436, No.\ AST-1516150,
No.\ ACI-1516125, No.\ PHY-1726215.
This work used the Extreme Science and Engineering
Discovery Environment (XSEDE) [allocation TG-PHY060027N], which is
supported by NSF grant No. ACI-1548562.
Computational resources were also provided by the NewHorizons,
BlueSky Clusters, and Green Prairies
at the Rochester Institute of Technology, which were
supported by NSF grants No.\ PHY-0722703, No.\ DMS-0820923, No.\
AST-1028087, No.\ PHY-1229173, and No.\ PHY-1726215.
Computational resources were also provided by the Blue Waters sustained-petascale computing NSF projects OAC-1811228, OAC-0832606, OAC-1238993, OAC- 1516247 and OAC-1515969, OAC-0725070 and by Frontera projects PHY-20010 and PHY-20007. Blue Waters is a joint effort of the University of Illinois at Urbana-Champaign and its National Center for Supercomputing Applications. Frontera is an NSF-funded petascale computing system at the Texas Advanced Computing Center (TACC).
\end{acknowledgments}

\bibliographystyle{apsrev4-1}
\bibliography{../../../../Bibtex/references}

\begin{thebibliography}{63}%
\makeatletter
\providecommand \@ifxundefined [1]{%
 \@ifx{#1\undefined}
}%
\providecommand \@ifnum [1]{%
 \ifnum #1\expandafter \@firstoftwo
 \else \expandafter \@secondoftwo
 \fi
}%
\providecommand \@ifx [1]{%
 \ifx #1\expandafter \@firstoftwo
 \else \expandafter \@secondoftwo
 \fi
}%
\providecommand \natexlab [1]{#1}%
\providecommand \enquote  [1]{``#1''}%
\providecommand \bibnamefont  [1]{#1}%
\providecommand \bibfnamefont [1]{#1}%
\providecommand \citenamefont [1]{#1}%
\providecommand \href@noop [0]{\@secondoftwo}%
\providecommand \href [0]{\begingroup \@sanitize@url \@href}%
\providecommand \@href[1]{\@@startlink{#1}\@@href}%
\providecommand \@@href[1]{\endgroup#1\@@endlink}%
\providecommand \@sanitize@url [0]{\catcode `\\12\catcode `\$12\catcode
  `\&12\catcode `\#12\catcode `\^12\catcode `\_12\catcode `\%12\relax}%
\providecommand \@@startlink[1]{}%
\providecommand \@@endlink[0]{}%
\providecommand \url  [0]{\begingroup\@sanitize@url \@url }%
\providecommand \@url [1]{\endgroup\@href {#1}{\urlprefix }}%
\providecommand \urlprefix  [0]{URL }%
\providecommand \Eprint [0]{\href }%
\providecommand \doibase [0]{https://doi.org/}%
\providecommand \selectlanguage [0]{\@gobble}%
\providecommand \bibinfo  [0]{\@secondoftwo}%
\providecommand \bibfield  [0]{\@secondoftwo}%
\providecommand \translation [1]{[#1]}%
\providecommand \BibitemOpen [0]{}%
\providecommand \bibitemStop [0]{}%
\providecommand \bibitemNoStop [0]{.\EOS\space}%
\providecommand \EOS [0]{\spacefactor3000\relax}%
\providecommand \BibitemShut  [1]{\csname bibitem#1\endcsname}%
\let\auto@bib@innerbib\@empty
%</preamble>
\bibitem [{\citenamefont {Campanelli}\ \emph
  {et~al.}(2007{\natexlab{a}})\citenamefont {Campanelli}, \citenamefont
  {Lousto}, \citenamefont {Zlochower},\ and\ \citenamefont
  {Merritt}}]{Campanelli:2007ew}%
  \BibitemOpen
  \bibfield  {author} {\bibinfo {author} {\bibfnamefont {M.}~\bibnamefont
  {Campanelli}}, \bibinfo {author} {\bibfnamefont {C.~O.}\ \bibnamefont
  {Lousto}}, \bibinfo {author} {\bibfnamefont {Y.}~\bibnamefont {Zlochower}},\
  and\ \bibinfo {author} {\bibfnamefont {D.}~\bibnamefont {Merritt}},\
  }\href@noop {} {\bibfield  {journal} {\bibinfo  {journal} {Astrophys. J.}\
  }\textbf {\bibinfo {volume} {659}},\ \bibinfo {pages} {L5} (\bibinfo {year}
  {2007}{\natexlab{a}})},\ \Eprint {https://arxiv.org/abs/gr-qc/0701164}
  {gr-qc/0701164} \BibitemShut {NoStop}%
%%CITATION = GR-QC/0701164;%%
\bibitem [{\citenamefont {Gonz\'alez}\ \emph {et~al.}(2007)\citenamefont
  {Gonz\'alez}, \citenamefont {Hannam}, \citenamefont {Sperhake}, \citenamefont
  {Brugmann},\ and\ \citenamefont {Husa}}]{Gonzalez:2007hi}%
  \BibitemOpen
  \bibfield  {author} {\bibinfo {author} {\bibfnamefont {J.~A.}\ \bibnamefont
  {Gonz\'alez}}, \bibinfo {author} {\bibfnamefont {M.~D.}\ \bibnamefont
  {Hannam}}, \bibinfo {author} {\bibfnamefont {U.}~\bibnamefont {Sperhake}},
  \bibinfo {author} {\bibfnamefont {B.}~\bibnamefont {Brugmann}},\ and\
  \bibinfo {author} {\bibfnamefont {S.}~\bibnamefont {Husa}},\ }\href@noop {}
  {\bibfield  {journal} {\bibinfo  {journal} {Phys. Rev. Lett.}\ }\textbf
  {\bibinfo {volume} {98}},\ \bibinfo {pages} {231101} (\bibinfo {year}
  {2007})},\ \Eprint {https://arxiv.org/abs/gr-qc/0702052} {gr-qc/0702052}
  \BibitemShut {NoStop}%
%%CITATION = GR-QC/0702052;%%
\bibitem [{\citenamefont {Campanelli}\ \emph
  {et~al.}(2007{\natexlab{b}})\citenamefont {Campanelli}, \citenamefont
  {Lousto}, \citenamefont {Zlochower},\ and\ \citenamefont
  {Merritt}}]{Campanelli:2007cga}%
  \BibitemOpen
  \bibfield  {author} {\bibinfo {author} {\bibfnamefont {M.}~\bibnamefont
  {Campanelli}}, \bibinfo {author} {\bibfnamefont {C.~O.}\ \bibnamefont
  {Lousto}}, \bibinfo {author} {\bibfnamefont {Y.}~\bibnamefont {Zlochower}},\
  and\ \bibinfo {author} {\bibfnamefont {D.}~\bibnamefont {Merritt}},\
  }\href@noop {} {\bibfield  {journal} {\bibinfo  {journal} {Phys. Rev. Lett.}\
  }\textbf {\bibinfo {volume} {98}},\ \bibinfo {pages} {231102} (\bibinfo
  {year} {2007}{\natexlab{b}})},\ \Eprint {https://arxiv.org/abs/gr-qc/0702133}
  {gr-qc/0702133} \BibitemShut {NoStop}%
%%CITATION = GR-QC/0702133;%%
\bibitem [{\citenamefont {Komossa}(2012)}]{Komossa:2012cy}%
  \BibitemOpen
  \bibfield  {author} {\bibinfo {author} {\bibfnamefont {S.}~\bibnamefont
  {Komossa}},\ }\href@noop {} {\bibfield  {journal} {\bibinfo  {journal} {Adv.
  Astron.}\ }\textbf {\bibinfo {volume} {2012}},\ \bibinfo {pages} {364973}
  (\bibinfo {year} {2012})},\ \Eprint {https://arxiv.org/abs/1202.1977}
  {arXiv:1202.1977 [astro-ph.CO]} \BibitemShut {NoStop}%
%%CITATION = ARXIV:1202.1977;%%
\bibitem [{\citenamefont {Chiaberge}\ \emph {et~al.}(2017)\citenamefont
  {Chiaberge} \emph {et~al.}}]{Chiaberge:2016eqf}%
  \BibitemOpen
  \bibfield  {author} {\bibinfo {author} {\bibfnamefont {M.}~\bibnamefont
  {Chiaberge}} \emph {et~al.},\ }\href
  {https://doi.org/10.1051/0004-6361/201629522} {\bibfield  {journal} {\bibinfo
   {journal} {Astron. Astrophys.}\ }\textbf {\bibinfo {volume} {600}},\
  \bibinfo {pages} {A57} (\bibinfo {year} {2017})},\ \Eprint
  {https://arxiv.org/abs/1611.05501} {arXiv:1611.05501 [astro-ph.GA]}
  \BibitemShut {NoStop}%
%%CITATION = ARXIV:1611.05501;%%
\bibitem [{\citenamefont {Lousto}\ \emph {et~al.}(2017)\citenamefont {Lousto},
  \citenamefont {Zlochower},\ and\ \citenamefont
  {Campanelli}}]{Lousto:2017uav}%
  \BibitemOpen
  \bibfield  {author} {\bibinfo {author} {\bibfnamefont {C.~O.}\ \bibnamefont
  {Lousto}}, \bibinfo {author} {\bibfnamefont {Y.}~\bibnamefont {Zlochower}},\
  and\ \bibinfo {author} {\bibfnamefont {M.}~\bibnamefont {Campanelli}},\
  }\href {https://doi.org/10.3847/2041-8213/aa733c} {\bibfield  {journal}
  {\bibinfo  {journal} {Astrophys. J.}\ }\textbf {\bibinfo {volume} {841}},\
  \bibinfo {pages} {L28} (\bibinfo {year} {2017})},\ \Eprint
  {https://arxiv.org/abs/1704.00809} {arXiv:1704.00809 [astro-ph.GA]}
  \BibitemShut {NoStop}%
%%CITATION = ARXIV:1704.00809;%%
\bibitem [{\citenamefont {Chiaberge}\ \emph {et~al.}(2018)\citenamefont
  {Chiaberge}, \citenamefont {Tremblay}, \citenamefont {Capetti},\ and\
  \citenamefont {Norman}}]{Chiaberge:2018lkg}%
  \BibitemOpen
  \bibfield  {author} {\bibinfo {author} {\bibfnamefont {M.}~\bibnamefont
  {Chiaberge}}, \bibinfo {author} {\bibfnamefont {G.~R.}\ \bibnamefont
  {Tremblay}}, \bibinfo {author} {\bibfnamefont {A.}~\bibnamefont {Capetti}},\
  and\ \bibinfo {author} {\bibfnamefont {C.}~\bibnamefont {Norman}},\ }\href
  {https://doi.org/10.3847/1538-4357/aac48b} {\bibfield  {journal} {\bibinfo
  {journal} {Astrophys. J.}\ }\textbf {\bibinfo {volume} {861}},\ \bibinfo
  {pages} {56} (\bibinfo {year} {2018})},\ \Eprint
  {https://arxiv.org/abs/1805.05860} {arXiv:1805.05860 [astro-ph.GA]}
  \BibitemShut {NoStop}%
%%CITATION = ARXIV:1805.05860;%%
\bibitem [{\citenamefont {Lousto}\ \emph
  {et~al.}(2010{\natexlab{a}})\citenamefont {Lousto}, \citenamefont {Nakano},
  \citenamefont {Zlochower},\ and\ \citenamefont {Campanelli}}]{Lousto:2009ka}%
  \BibitemOpen
  \bibfield  {author} {\bibinfo {author} {\bibfnamefont {C.~O.}\ \bibnamefont
  {Lousto}}, \bibinfo {author} {\bibfnamefont {H.}~\bibnamefont {Nakano}},
  \bibinfo {author} {\bibfnamefont {Y.}~\bibnamefont {Zlochower}},\ and\
  \bibinfo {author} {\bibfnamefont {M.}~\bibnamefont {Campanelli}},\
  }\href@noop {} {\bibfield  {journal} {\bibinfo  {journal} {Phys. Rev.}\
  }\textbf {\bibinfo {volume} {D81}},\ \bibinfo {pages} {084023} (\bibinfo
  {year} {2010}{\natexlab{a}})},\ \Eprint {https://arxiv.org/abs/0910.3197}
  {arXiv:0910.3197 [gr-qc]} \BibitemShut {NoStop}%
%%CITATION = 0910.3197;%%
\bibitem [{\citenamefont {Fishbach}\ \emph {et~al.}(2017)\citenamefont
  {Fishbach}, \citenamefont {Holz},\ and\ \citenamefont
  {Farr}}]{Fishbach:2017dwv}%
  \BibitemOpen
  \bibfield  {author} {\bibinfo {author} {\bibfnamefont {M.}~\bibnamefont
  {Fishbach}}, \bibinfo {author} {\bibfnamefont {D.~E.}\ \bibnamefont {Holz}},\
  and\ \bibinfo {author} {\bibfnamefont {B.}~\bibnamefont {Farr}},\ }\href
  {https://doi.org/10.3847/2041-8213/aa7045} {\bibfield  {journal} {\bibinfo
  {journal} {Astrophys. J.}\ }\textbf {\bibinfo {volume} {840}},\ \bibinfo
  {pages} {L24} (\bibinfo {year} {2017})},\ \Eprint
  {https://arxiv.org/abs/1703.06869} {arXiv:1703.06869 [astro-ph.HE]}
  \BibitemShut {NoStop}%
%%CITATION = ARXIV:1703.06869;%%
\bibitem [{\citenamefont {Lousto}\ \emph {et~al.}(2012)\citenamefont {Lousto},
  \citenamefont {Zlochower}, \citenamefont {Dotti},\ and\ \citenamefont
  {Volonteri}}]{Lousto:2012su}%
  \BibitemOpen
  \bibfield  {author} {\bibinfo {author} {\bibfnamefont {C.~O.}\ \bibnamefont
  {Lousto}}, \bibinfo {author} {\bibfnamefont {Y.}~\bibnamefont {Zlochower}},
  \bibinfo {author} {\bibfnamefont {M.}~\bibnamefont {Dotti}},\ and\ \bibinfo
  {author} {\bibfnamefont {M.}~\bibnamefont {Volonteri}},\ }\href@noop {}
  {\bibfield  {journal} {\bibinfo  {journal} {Phys. Rev.}\ }\textbf {\bibinfo
  {volume} {D85}},\ \bibinfo {pages} {084015} (\bibinfo {year} {2012})},\
  \Eprint {https://arxiv.org/abs/1201.1923} {arXiv:1201.1923 [gr-qc]}
  \BibitemShut {NoStop}%
%%CITATION = ARXIV:1201.1923;%%
\bibitem [{\citenamefont {Rodriguez}\ \emph {et~al.}(2018)\citenamefont
  {Rodriguez}, \citenamefont {Amaro-Seoane}, \citenamefont {Chatterjee},\ and\
  \citenamefont {Rasio}}]{Rodriguez:2019huv}%
  \BibitemOpen
  \bibfield  {author} {\bibinfo {author} {\bibfnamefont {C.~L.}\ \bibnamefont
  {Rodriguez}}, \bibinfo {author} {\bibfnamefont {P.}~\bibnamefont
  {Amaro-Seoane}}, \bibinfo {author} {\bibfnamefont {S.}~\bibnamefont
  {Chatterjee}},\ and\ \bibinfo {author} {\bibfnamefont {F.~A.}\ \bibnamefont
  {Rasio}},\ }\href {https://doi.org/10.1103/PhysRevLett.120.151101} {\bibfield
   {journal} {\bibinfo  {journal} {Phys. Rev. Lett.}\ }\textbf {\bibinfo
  {volume} {120}},\ \bibinfo {pages} {151101} (\bibinfo {year} {2018})},\
  \Eprint {https://arxiv.org/abs/1712.04937} {arXiv:1712.04937 [astro-ph.HE]}
  \BibitemShut {NoStop}%
%%CITATION = ARXIV:1712.04937;%%
\bibitem [{\citenamefont {Lousto}\ and\ \citenamefont
  {Healy}(2019)}]{Lousto:2019lyf}%
  \BibitemOpen
  \bibfield  {author} {\bibinfo {author} {\bibfnamefont {C.~O.}\ \bibnamefont
  {Lousto}}\ and\ \bibinfo {author} {\bibfnamefont {J.}~\bibnamefont {Healy}},\
  }\href {https://doi.org/10.1103/PhysRevD.100.104039} {\bibfield  {journal}
  {\bibinfo  {journal} {Phys. Rev.}\ }\textbf {\bibinfo {volume} {D100}},\
  \bibinfo {pages} {104039} (\bibinfo {year} {2019})},\ \Eprint
  {https://arxiv.org/abs/1908.04382} {arXiv:1908.04382 [gr-qc]} \BibitemShut
  {NoStop}%
%%CITATION = ARXIV:1908.04382;%%
\bibitem [{\citenamefont {Sesana}\ \emph {et~al.}(2009)\citenamefont {Sesana},
  \citenamefont {Volonteri},\ and\ \citenamefont {Haardt}}]{Sesana:2008ur}%
  \BibitemOpen
  \bibfield  {author} {\bibinfo {author} {\bibfnamefont {A.}~\bibnamefont
  {Sesana}}, \bibinfo {author} {\bibfnamefont {M.}~\bibnamefont {Volonteri}},\
  and\ \bibinfo {author} {\bibfnamefont {F.}~\bibnamefont {Haardt}},\
  }\bibfield  {booktitle} {\emph {\bibinfo {booktitle} {{Laser Interferometer
  Space Antenna. Proceedings, 7th international LISA Symposium, Barcelona,
  Spain, June 16-20, 2008}}},\ }\href
  {https://doi.org/10.1088/0264-9381/26/9/094033} {\bibfield  {journal}
  {\bibinfo  {journal} {Class. Quant. Grav.}\ }\textbf {\bibinfo {volume}
  {26}},\ \bibinfo {pages} {094033} (\bibinfo {year} {2009})},\ \Eprint
  {https://arxiv.org/abs/0810.5554} {arXiv:0810.5554 [astro-ph]} \BibitemShut
  {NoStop}%
%%CITATION = ARXIV:0810.5554;%%
\bibitem [{\citenamefont {Gerosa}\ and\ \citenamefont
  {Moore}(2016)}]{Gerosa:2016vip}%
  \BibitemOpen
  \bibfield  {author} {\bibinfo {author} {\bibfnamefont {D.}~\bibnamefont
  {Gerosa}}\ and\ \bibinfo {author} {\bibfnamefont {C.~J.}\ \bibnamefont
  {Moore}},\ }\href {https://doi.org/10.1103/PhysRevLett.117.011101} {\bibfield
   {journal} {\bibinfo  {journal} {Phys. Rev. Lett.}\ }\textbf {\bibinfo
  {volume} {117}},\ \bibinfo {pages} {011101} (\bibinfo {year} {2016})},\
  \Eprint {https://arxiv.org/abs/1606.04226} {arXiv:1606.04226 [gr-qc]}
  \BibitemShut {NoStop}%
%%CITATION = ARXIV:1606.04226;%%
\bibitem [{\citenamefont {Abbott}\ \emph
  {et~al.}(2016{\natexlab{a}})\citenamefont {Abbott} \emph
  {et~al.}}]{Abbott:2016apu}%
  \BibitemOpen
  \bibfield  {author} {\bibinfo {author} {\bibfnamefont {B.~P.}\ \bibnamefont
  {Abbott}} \emph {et~al.} (\bibinfo {collaboration} {Virgo, LIGO
  Scientific}),\ }\href {https://doi.org/10.1103/PhysRevD.94.064035} {\bibfield
   {journal} {\bibinfo  {journal} {Phys. Rev.}\ }\textbf {\bibinfo {volume}
  {D94}},\ \bibinfo {pages} {064035} (\bibinfo {year} {2016}{\natexlab{a}})},\
  \Eprint {https://arxiv.org/abs/1606.01262} {arXiv:1606.01262 [gr-qc]}
  \BibitemShut {NoStop}%
%%CITATION = ARXIV:1606.01262;%%
\bibitem [{\citenamefont {Lovelace}\ \emph {et~al.}(2016)\citenamefont
  {Lovelace} \emph {et~al.}}]{Lovelace:2016uwp}%
  \BibitemOpen
  \bibfield  {author} {\bibinfo {author} {\bibfnamefont {G.}~\bibnamefont
  {Lovelace}} \emph {et~al.},\ }\href
  {https://doi.org/10.1088/0264-9381/33/24/244002} {\bibfield  {journal}
  {\bibinfo  {journal} {Class. Quant. Grav.}\ }\textbf {\bibinfo {volume}
  {33}},\ \bibinfo {pages} {244002} (\bibinfo {year} {2016})},\ \Eprint
  {https://arxiv.org/abs/1607.05377} {arXiv:1607.05377 [gr-qc]} \BibitemShut
  {NoStop}%
%%CITATION = ARXIV:1607.05377;%%
\bibitem [{\citenamefont {Healy}\ \emph {et~al.}(2019)\citenamefont {Healy},
  \citenamefont {Lousto}, \citenamefont {Lange}, \citenamefont {O'Shaughnessy},
  \citenamefont {Zlochower},\ and\ \citenamefont {Campanelli}}]{Healy:2019jyf}%
  \BibitemOpen
  \bibfield  {author} {\bibinfo {author} {\bibfnamefont {J.}~\bibnamefont
  {Healy}}, \bibinfo {author} {\bibfnamefont {C.~O.}\ \bibnamefont {Lousto}},
  \bibinfo {author} {\bibfnamefont {J.}~\bibnamefont {Lange}}, \bibinfo
  {author} {\bibfnamefont {R.}~\bibnamefont {O'Shaughnessy}}, \bibinfo {author}
  {\bibfnamefont {Y.}~\bibnamefont {Zlochower}},\ and\ \bibinfo {author}
  {\bibfnamefont {M.}~\bibnamefont {Campanelli}},\ }\href
  {https://doi.org/10.1103/PhysRevD.100.024021} {\bibfield  {journal} {\bibinfo
   {journal} {Phys. Rev.}\ }\textbf {\bibinfo {volume} {D100}},\ \bibinfo
  {pages} {024021} (\bibinfo {year} {2019})},\ \Eprint
  {https://arxiv.org/abs/1901.02553} {arXiv:1901.02553 [gr-qc]} \BibitemShut
  {NoStop}%
%%CITATION = ARXIV:1901.02553;%%
\bibitem [{\citenamefont {Healy}\ \emph
  {et~al.}(2018{\natexlab{a}})\citenamefont {Healy} \emph
  {et~al.}}]{Heal:2017abq}%
  \BibitemOpen
  \bibfield  {author} {\bibinfo {author} {\bibfnamefont {J.}~\bibnamefont
  {Healy}} \emph {et~al.},\ }\href {https://doi.org/10.1103/PhysRevD.97.064027}
  {\bibfield  {journal} {\bibinfo  {journal} {Phys. Rev.}\ }\textbf {\bibinfo
  {volume} {D97}},\ \bibinfo {pages} {064027} (\bibinfo {year}
  {2018}{\natexlab{a}})},\ \Eprint {https://arxiv.org/abs/1712.05836}
  {arXiv:1712.05836 [gr-qc]} \BibitemShut {NoStop}%
%%CITATION = ARXIV:1712.05836;%%
\bibitem [{\citenamefont {Abbott}\ \emph
  {et~al.}(2019{\natexlab{a}})\citenamefont {Abbott} \emph
  {et~al.}}]{LIGOScientific:2018mvr}%
  \BibitemOpen
  \bibfield  {author} {\bibinfo {author} {\bibfnamefont {B.~P.}\ \bibnamefont
  {Abbott}} \emph {et~al.} (\bibinfo {collaboration} {LIGO Scientific,
  Virgo}),\ }\href {https://doi.org/10.1103/PhysRevX.9.031040} {\bibfield
  {journal} {\bibinfo  {journal} {Phys. Rev.}\ }\textbf {\bibinfo {volume}
  {X9}},\ \bibinfo {pages} {031040} (\bibinfo {year} {2019}{\natexlab{a}})},\
  \Eprint {https://arxiv.org/abs/1811.12907} {arXiv:1811.12907 [astro-ph.HE]}
  \BibitemShut {NoStop}%
%%CITATION = ARXIV:1811.12907;%%
\bibitem [{\citenamefont {Abbott}\ \emph
  {et~al.}(2016{\natexlab{b}})\citenamefont {Abbott} \emph
  {et~al.}}]{TheLIGOScientific:2016src}%
  \BibitemOpen
  \bibfield  {author} {\bibinfo {author} {\bibfnamefont {B.~P.}\ \bibnamefont
  {Abbott}} \emph {et~al.} (\bibinfo {collaboration} {Virgo, LIGO
  Scientific}),\ }\href {https://doi.org/10.1103/PhysRevLett.116.221101}
  {\bibfield  {journal} {\bibinfo  {journal} {Phys. Rev. Lett.}\ }\textbf
  {\bibinfo {volume} {116}},\ \bibinfo {pages} {221101} (\bibinfo {year}
  {2016}{\natexlab{b}})},\ \Eprint {https://arxiv.org/abs/1602.03841}
  {arXiv:1602.03841 [gr-qc]} \BibitemShut {NoStop}%
%%CITATION = ARXIV:1602.03841;%%
\bibitem [{\citenamefont {Abbott}\ \emph
  {et~al.}(2019{\natexlab{b}})\citenamefont {Abbott} \emph
  {et~al.}}]{LIGOScientific:2019fpa}%
  \BibitemOpen
  \bibfield  {author} {\bibinfo {author} {\bibfnamefont {B.~P.}\ \bibnamefont
  {Abbott}} \emph {et~al.} (\bibinfo {collaboration} {LIGO Scientific,
  Virgo}),\ }\href {https://doi.org/10.1103/PhysRevD.100.104036} {\bibfield
  {journal} {\bibinfo  {journal} {Phys. Rev.}\ }\textbf {\bibinfo {volume}
  {D100}},\ \bibinfo {pages} {104036} (\bibinfo {year} {2019}{\natexlab{b}})},\
  \Eprint {https://arxiv.org/abs/1903.04467} {arXiv:1903.04467 [gr-qc]}
  \BibitemShut {NoStop}%
%%CITATION = ARXIV:1903.04467;%%
\bibitem [{\citenamefont {Echeverr\'{\i}a}(1989)}]{Echeverria89}%
  \BibitemOpen
  \bibfield  {author} {\bibinfo {author} {\bibfnamefont {F.}~\bibnamefont
  {Echeverr\'{\i}a}},\ }\href@noop {} {\bibfield  {journal} {\bibinfo
  {journal} {Phys. Rev. D}\ }\textbf {\bibinfo {volume} {40}},\ \bibinfo
  {pages} {3194} (\bibinfo {year} {1989})}\BibitemShut {NoStop}%
\bibitem [{\citenamefont {Berti}\ \emph {et~al.}(2006)\citenamefont {Berti},
  \citenamefont {Cardoso},\ and\ \citenamefont {Will}}]{Berti:2005ys}%
  \BibitemOpen
  \bibfield  {author} {\bibinfo {author} {\bibfnamefont {E.}~\bibnamefont
  {Berti}}, \bibinfo {author} {\bibfnamefont {V.}~\bibnamefont {Cardoso}},\
  and\ \bibinfo {author} {\bibfnamefont {C.~M.}\ \bibnamefont {Will}},\ }\href
  {https://doi.org/10.1103/PhysRevD.73.064030} {\bibfield  {journal} {\bibinfo
  {journal} {Phys. Rev.}\ }\textbf {\bibinfo {volume} {D73}},\ \bibinfo {pages}
  {064030} (\bibinfo {year} {2006})},\ \Eprint
  {https://arxiv.org/abs/gr-qc/0512160} {arXiv:gr-qc/0512160 [gr-qc]}
  \BibitemShut {NoStop}%
%%CITATION = GR-QC/0512160;%%
\bibitem [{\citenamefont {Berti}\ \emph {et~al.}(2009)\citenamefont {Berti},
  \citenamefont {Cardoso},\ and\ \citenamefont {Starinets}}]{Berti:2009kk}%
  \BibitemOpen
  \bibfield  {author} {\bibinfo {author} {\bibfnamefont {E.}~\bibnamefont
  {Berti}}, \bibinfo {author} {\bibfnamefont {V.}~\bibnamefont {Cardoso}},\
  and\ \bibinfo {author} {\bibfnamefont {A.~O.}\ \bibnamefont {Starinets}},\
  }\href {https://doi.org/10.1088/0264-9381/26/16/163001} {\bibfield  {journal}
  {\bibinfo  {journal} {Class. Quant. Grav.}\ }\textbf {\bibinfo {volume}
  {26}},\ \bibinfo {pages} {163001} (\bibinfo {year} {2009})},\ \Eprint
  {https://arxiv.org/abs/0905.2975} {arXiv:0905.2975 [gr-qc]} \BibitemShut
  {NoStop}%
%%CITATION = 0905.2975;%%
\bibitem [{\citenamefont {Dreyer}\ \emph
  {et~al.}(2003{\natexlab{a}})\citenamefont {Dreyer}, \citenamefont {Krishnan},
  \citenamefont {Shoemaker},\ and\ \citenamefont {Schnetter}}]{Dreyer:2002mx}%
  \BibitemOpen
  \bibfield  {author} {\bibinfo {author} {\bibfnamefont {O.}~\bibnamefont
  {Dreyer}}, \bibinfo {author} {\bibfnamefont {B.}~\bibnamefont {Krishnan}},
  \bibinfo {author} {\bibfnamefont {D.}~\bibnamefont {Shoemaker}},\ and\
  \bibinfo {author} {\bibfnamefont {E.}~\bibnamefont {Schnetter}},\ }\href@noop
  {} {\bibfield  {journal} {\bibinfo  {journal} {Phys. Rev.}\ }\textbf
  {\bibinfo {volume} {D67}},\ \bibinfo {pages} {024018} (\bibinfo {year}
  {2003}{\natexlab{a}})},\ \Eprint {https://arxiv.org/abs/gr-qc/0206008}
  {gr-qc/0206008} \BibitemShut {NoStop}%
%%CITATION = GR-QC/0206008;%%
\bibitem [{\citenamefont {Dain}\ \emph {et~al.}(2008)\citenamefont {Dain},
  \citenamefont {Lousto},\ and\ \citenamefont {Zlochower}}]{Dain:2008ck}%
  \BibitemOpen
  \bibfield  {author} {\bibinfo {author} {\bibfnamefont {S.}~\bibnamefont
  {Dain}}, \bibinfo {author} {\bibfnamefont {C.~O.}\ \bibnamefont {Lousto}},\
  and\ \bibinfo {author} {\bibfnamefont {Y.}~\bibnamefont {Zlochower}},\ }\href
  {https://doi.org/10.1103/PhysRevD.78.024039} {\bibfield  {journal} {\bibinfo
  {journal} {Phys. Rev.}\ }\textbf {\bibinfo {volume} {D78}},\ \bibinfo {pages}
  {024039} (\bibinfo {year} {2008})},\ \Eprint
  {https://arxiv.org/abs/0803.0351} {arXiv:0803.0351 [gr-qc]} \BibitemShut
  {NoStop}%
%%CITATION = 0803.0351;%%
\bibitem [{\citenamefont {Healy}\ and\ \citenamefont
  {Lousto}(2018)}]{Healy:2018swt}%
  \BibitemOpen
  \bibfield  {author} {\bibinfo {author} {\bibfnamefont {J.}~\bibnamefont
  {Healy}}\ and\ \bibinfo {author} {\bibfnamefont {C.~O.}\ \bibnamefont
  {Lousto}},\ }\href {https://doi.org/10.1103/PhysRevD.97.084002} {\bibfield
  {journal} {\bibinfo  {journal} {Phys. Rev.}\ }\textbf {\bibinfo {volume}
  {D97}},\ \bibinfo {pages} {084002} (\bibinfo {year} {2018})},\ \Eprint
  {https://arxiv.org/abs/1801.08162} {arXiv:1801.08162 [gr-qc]} \BibitemShut
  {NoStop}%
%%CITATION = ARXIV:1801.08162;%%
\bibitem [{\citenamefont {Krishnan}\ \emph {et~al.}(2007)\citenamefont
  {Krishnan}, \citenamefont {Lousto},\ and\ \citenamefont
  {Zlochower}}]{Krishnan:2007pu}%
  \BibitemOpen
  \bibfield  {author} {\bibinfo {author} {\bibfnamefont {B.}~\bibnamefont
  {Krishnan}}, \bibinfo {author} {\bibfnamefont {C.~O.}\ \bibnamefont
  {Lousto}},\ and\ \bibinfo {author} {\bibfnamefont {Y.}~\bibnamefont
  {Zlochower}},\ }\href {https://doi.org/10.1103/PhysRevD.76.081501} {\bibfield
   {journal} {\bibinfo  {journal} {Phys. Rev.}\ }\textbf {\bibinfo {volume}
  {D76}},\ \bibinfo {pages} {081501} (\bibinfo {year} {2007})},\ \Eprint
  {https://arxiv.org/abs/0707.0876} {arXiv:0707.0876 [gr-qc]} \BibitemShut
  {NoStop}%
%%CITATION = 0707.0876;%%
\bibitem [{\citenamefont {van Meter}\ \emph {et~al.}(2006)\citenamefont {van
  Meter}, \citenamefont {Baker}, \citenamefont {Koppitz},\ and\ \citenamefont
  {Choi}}]{vanMeter:2006vi}%
  \BibitemOpen
  \bibfield  {author} {\bibinfo {author} {\bibfnamefont {J.~R.}\ \bibnamefont
  {van Meter}}, \bibinfo {author} {\bibfnamefont {J.~G.}\ \bibnamefont
  {Baker}}, \bibinfo {author} {\bibfnamefont {M.}~\bibnamefont {Koppitz}},\
  and\ \bibinfo {author} {\bibfnamefont {D.-I.}\ \bibnamefont {Choi}},\
  }\href@noop {} {\bibfield  {journal} {\bibinfo  {journal} {Phys. Rev.}\
  }\textbf {\bibinfo {volume} {D73}},\ \bibinfo {pages} {124011} (\bibinfo
  {year} {2006})},\ \Eprint {https://arxiv.org/abs/gr-qc/0605030}
  {gr-qc/0605030} \BibitemShut {NoStop}%
%%CITATION = GR-QC 0605030;%%
\bibitem [{\citenamefont {Campanelli}\ \emph {et~al.}(2006)\citenamefont
  {Campanelli}, \citenamefont {Lousto}, \citenamefont {Marronetti},\ and\
  \citenamefont {Zlochower}}]{Campanelli:2005dd}%
  \BibitemOpen
  \bibfield  {author} {\bibinfo {author} {\bibfnamefont {M.}~\bibnamefont
  {Campanelli}}, \bibinfo {author} {\bibfnamefont {C.~O.}\ \bibnamefont
  {Lousto}}, \bibinfo {author} {\bibfnamefont {P.}~\bibnamefont {Marronetti}},\
  and\ \bibinfo {author} {\bibfnamefont {Y.}~\bibnamefont {Zlochower}},\
  }\href@noop {} {\bibfield  {journal} {\bibinfo  {journal} {Phys. Rev. Lett.}\
  }\textbf {\bibinfo {volume} {96}},\ \bibinfo {pages} {111101} (\bibinfo
  {year} {2006})},\ \Eprint {https://arxiv.org/abs/gr-qc/0511048}
  {gr-qc/0511048} \BibitemShut {NoStop}%
%%CITATION = GR-QC 0511048;%%
\bibitem [{\citenamefont {Nakamura}\ \emph {et~al.}(1987)\citenamefont
  {Nakamura}, \citenamefont {Oohara},\ and\ \citenamefont
  {Kojima}}]{Nakamura87}%
  \BibitemOpen
  \bibfield  {author} {\bibinfo {author} {\bibfnamefont {T.}~\bibnamefont
  {Nakamura}}, \bibinfo {author} {\bibfnamefont {K.}~\bibnamefont {Oohara}},\
  and\ \bibinfo {author} {\bibfnamefont {Y.}~\bibnamefont {Kojima}},\
  }\href@noop {} {\bibfield  {journal} {\bibinfo  {journal} {Prog. Theor. Phys.
  Suppl.}\ }\textbf {\bibinfo {volume} {90}},\ \bibinfo {pages} {1} (\bibinfo
  {year} {1987})}\BibitemShut {NoStop}%
\bibitem [{\citenamefont {Shibata}\ and\ \citenamefont
  {Nakamura}(1995)}]{Shibata95}%
  \BibitemOpen
  \bibfield  {author} {\bibinfo {author} {\bibfnamefont {M.}~\bibnamefont
  {Shibata}}\ and\ \bibinfo {author} {\bibfnamefont {T.}~\bibnamefont
  {Nakamura}},\ }\href@noop {} {\bibfield  {journal} {\bibinfo  {journal}
  {Phys. Rev.}\ }\textbf {\bibinfo {volume} {D52}},\ \bibinfo {pages} {5428}
  (\bibinfo {year} {1995})}\BibitemShut {NoStop}%
\bibitem [{\citenamefont {Baumgarte}\ and\ \citenamefont
  {Shapiro}(1998)}]{Baumgarte99}%
  \BibitemOpen
  \bibfield  {author} {\bibinfo {author} {\bibfnamefont {T.~W.}\ \bibnamefont
  {Baumgarte}}\ and\ \bibinfo {author} {\bibfnamefont {S.~L.}\ \bibnamefont
  {Shapiro}},\ }\href@noop {} {\bibfield  {journal} {\bibinfo  {journal} {Phys.
  Rev.}\ }\textbf {\bibinfo {volume} {D59}},\ \bibinfo {pages} {024007}
  (\bibinfo {year} {1998})},\ \Eprint {https://arxiv.org/abs/gr-qc/9810065}
  {gr-qc/9810065} \BibitemShut {NoStop}%
\bibitem [{\citenamefont {Alcubierre}\ \emph {et~al.}(2003)\citenamefont
  {Alcubierre}, \citenamefont {Br\"ugmann}, \citenamefont {Diener},
  \citenamefont {Koppitz}, \citenamefont {Pollney}, \citenamefont {Seidel},\
  and\ \citenamefont {Takahashi}}]{Alcubierre02a}%
  \BibitemOpen
  \bibfield  {author} {\bibinfo {author} {\bibfnamefont {M.}~\bibnamefont
  {Alcubierre}}, \bibinfo {author} {\bibfnamefont {B.}~\bibnamefont
  {Br\"ugmann}}, \bibinfo {author} {\bibfnamefont {P.}~\bibnamefont {Diener}},
  \bibinfo {author} {\bibfnamefont {M.}~\bibnamefont {Koppitz}}, \bibinfo
  {author} {\bibfnamefont {D.}~\bibnamefont {Pollney}}, \bibinfo {author}
  {\bibfnamefont {E.}~\bibnamefont {Seidel}},\ and\ \bibinfo {author}
  {\bibfnamefont {R.}~\bibnamefont {Takahashi}},\ }\href@noop {} {\bibfield
  {journal} {\bibinfo  {journal} {Phys. Rev.}\ }\textbf {\bibinfo {volume}
  {D67}},\ \bibinfo {pages} {084023} (\bibinfo {year} {2003})},\ \Eprint
  {https://arxiv.org/abs/gr-qc/0206072} {gr-qc/0206072} \BibitemShut {NoStop}%
%%CITATION = GR-QC 0206072;%%
\bibitem [{\citenamefont {Baker}\ \emph {et~al.}(2006)\citenamefont {Baker},
  \citenamefont {Centrella}, \citenamefont {Choi}, \citenamefont {Koppitz},\
  and\ \citenamefont {van Meter}}]{Baker:2005vv}%
  \BibitemOpen
  \bibfield  {author} {\bibinfo {author} {\bibfnamefont {J.~G.}\ \bibnamefont
  {Baker}}, \bibinfo {author} {\bibfnamefont {J.}~\bibnamefont {Centrella}},
  \bibinfo {author} {\bibfnamefont {D.-I.}\ \bibnamefont {Choi}}, \bibinfo
  {author} {\bibfnamefont {M.}~\bibnamefont {Koppitz}},\ and\ \bibinfo {author}
  {\bibfnamefont {J.}~\bibnamefont {van Meter}},\ }\href@noop {} {\bibfield
  {journal} {\bibinfo  {journal} {Phys. Rev. Lett.}\ }\textbf {\bibinfo
  {volume} {96}},\ \bibinfo {pages} {111102} (\bibinfo {year} {2006})},\
  \Eprint {https://arxiv.org/abs/gr-qc/0511103} {gr-qc/0511103} \BibitemShut
  {NoStop}%
%%CITATION = GR-QC 0511103;%%
\bibitem [{\citenamefont {Lousto}\ and\ \citenamefont
  {Zlochower}(2008)}]{Lousto:2007db}%
  \BibitemOpen
  \bibfield  {author} {\bibinfo {author} {\bibfnamefont {C.~O.}\ \bibnamefont
  {Lousto}}\ and\ \bibinfo {author} {\bibfnamefont {Y.}~\bibnamefont
  {Zlochower}},\ }\href {https://doi.org/10.1103/PhysRevD.77.044028} {\bibfield
   {journal} {\bibinfo  {journal} {Phys. Rev.}\ }\textbf {\bibinfo {volume}
  {D77}},\ \bibinfo {pages} {044028} (\bibinfo {year} {2008})},\ \Eprint
  {https://arxiv.org/abs/0708.4048} {arXiv:0708.4048 [gr-qc]} \BibitemShut
  {NoStop}%
%%CITATION = 0708.4048;%%
\bibitem [{\citenamefont {Ansorg}\ \emph {et~al.}(2004)\citenamefont {Ansorg},
  \citenamefont {Br\"ugmann},\ and\ \citenamefont {Tichy}}]{Ansorg:2004ds}%
  \BibitemOpen
  \bibfield  {author} {\bibinfo {author} {\bibfnamefont {M.}~\bibnamefont
  {Ansorg}}, \bibinfo {author} {\bibfnamefont {B.}~\bibnamefont {Br\"ugmann}},\
  and\ \bibinfo {author} {\bibfnamefont {W.}~\bibnamefont {Tichy}},\
  }\href@noop {} {\bibfield  {journal} {\bibinfo  {journal} {Phys. Rev.}\
  }\textbf {\bibinfo {volume} {D70}},\ \bibinfo {pages} {064011} (\bibinfo
  {year} {2004})},\ \Eprint {https://arxiv.org/abs/gr-qc/0404056}
  {gr-qc/0404056} \BibitemShut {NoStop}%
%%CITATION = GR-QC 0404056;%%
\bibitem [{\citenamefont {Zlochower}\ \emph {et~al.}(2005)\citenamefont
  {Zlochower}, \citenamefont {Baker}, \citenamefont {Campanelli},\ and\
  \citenamefont {Lousto}}]{Zlochower:2005bj}%
  \BibitemOpen
  \bibfield  {author} {\bibinfo {author} {\bibfnamefont {Y.}~\bibnamefont
  {Zlochower}}, \bibinfo {author} {\bibfnamefont {J.~G.}\ \bibnamefont
  {Baker}}, \bibinfo {author} {\bibfnamefont {M.}~\bibnamefont {Campanelli}},\
  and\ \bibinfo {author} {\bibfnamefont {C.~O.}\ \bibnamefont {Lousto}},\
  }\href {https://doi.org/10.1103/PhysRevD.72.024021} {\bibfield  {journal}
  {\bibinfo  {journal} {Phys. Rev.}\ }\textbf {\bibinfo {volume} {D72}},\
  \bibinfo {pages} {024021} (\bibinfo {year} {2005})},\ \Eprint
  {https://arxiv.org/abs/gr-qc/0505055} {arXiv:gr-qc/0505055} \BibitemShut
  {NoStop}%
%%CITATION = GR-QC/0505055;%%
\bibitem [{\citenamefont {Schnetter}\ \emph {et~al.}(2004)\citenamefont
  {Schnetter}, \citenamefont {Hawley},\ and\ \citenamefont
  {Hawke}}]{Schnetter-etal-03b}%
  \BibitemOpen
  \bibfield  {author} {\bibinfo {author} {\bibfnamefont {E.}~\bibnamefont
  {Schnetter}}, \bibinfo {author} {\bibfnamefont {S.~H.}\ \bibnamefont
  {Hawley}},\ and\ \bibinfo {author} {\bibfnamefont {I.}~\bibnamefont
  {Hawke}},\ }\href@noop {} {\bibfield  {journal} {\bibinfo  {journal} {Class.
  Quant. Grav.}\ }\textbf {\bibinfo {volume} {21}},\ \bibinfo {pages} {1465}
  (\bibinfo {year} {2004})},\ \Eprint {https://arxiv.org/abs/gr-qc/0310042}
  {gr-qc/0310042} \BibitemShut {NoStop}%
\bibitem [{car()}]{carpet_web}%
  \BibitemOpen
  \href@noop {} {}\bibinfo {note} {Carpet - adaptive mesh refinement for the
  cactus framework:\\{\tt https://carpetcode.org}}\BibitemShut {NoStop}%
\bibitem [{\citenamefont {Thornburg}(2004)}]{Thornburg2003:AH-finding}%
  \BibitemOpen
  \bibfield  {author} {\bibinfo {author} {\bibfnamefont {J.}~\bibnamefont
  {Thornburg}},\ }\href {https://doi.org/10.1088/0264-9381/21/2/026} {\bibfield
   {journal} {\bibinfo  {journal} {Class. Quant. Grav.}\ }\textbf {\bibinfo
  {volume} {21}},\ \bibinfo {pages} {743} (\bibinfo {year} {2004})},\ \Eprint
  {https://arxiv.org/abs/gr-qc/0306056} {gr-qc/0306056} \BibitemShut {NoStop}%
\bibitem [{\citenamefont {Dreyer}\ \emph
  {et~al.}(2003{\natexlab{b}})\citenamefont {Dreyer}, \citenamefont {Krishnan},
  \citenamefont {Shoemaker},\ and\ \citenamefont {Schnetter}}]{Dreyer02a}%
  \BibitemOpen
  \bibfield  {author} {\bibinfo {author} {\bibfnamefont {O.}~\bibnamefont
  {Dreyer}}, \bibinfo {author} {\bibfnamefont {B.}~\bibnamefont {Krishnan}},
  \bibinfo {author} {\bibfnamefont {D.}~\bibnamefont {Shoemaker}},\ and\
  \bibinfo {author} {\bibfnamefont {E.}~\bibnamefont {Schnetter}},\ }\href@noop
  {} {\bibfield  {journal} {\bibinfo  {journal} {Phys. Rev.}\ }\textbf
  {\bibinfo {volume} {D67}},\ \bibinfo {pages} {024018} (\bibinfo {year}
  {2003}{\natexlab{b}})},\ \Eprint {https://arxiv.org/abs/gr-qc/0206008}
  {gr-qc/0206008} \BibitemShut {NoStop}%
\bibitem [{\citenamefont {Campanelli}\ \emph
  {et~al.}(2007{\natexlab{c}})\citenamefont {Campanelli}, \citenamefont
  {Lousto}, \citenamefont {Zlochower}, \citenamefont {Krishnan},\ and\
  \citenamefont {Merritt}}]{Campanelli:2006fy}%
  \BibitemOpen
  \bibfield  {author} {\bibinfo {author} {\bibfnamefont {M.}~\bibnamefont
  {Campanelli}}, \bibinfo {author} {\bibfnamefont {C.~O.}\ \bibnamefont
  {Lousto}}, \bibinfo {author} {\bibfnamefont {Y.}~\bibnamefont {Zlochower}},
  \bibinfo {author} {\bibfnamefont {B.}~\bibnamefont {Krishnan}},\ and\
  \bibinfo {author} {\bibfnamefont {D.}~\bibnamefont {Merritt}},\ }\href@noop
  {} {\bibfield  {journal} {\bibinfo  {journal} {Phys. Rev.}\ }\textbf
  {\bibinfo {volume} {D75}},\ \bibinfo {pages} {064030} (\bibinfo {year}
  {2007}{\natexlab{c}})},\ \Eprint {https://arxiv.org/abs/gr-qc/0612076}
  {gr-qc/0612076} \BibitemShut {NoStop}%
%%CITATION = GR-QC/0612076;%%
\bibitem [{\citenamefont {Campanelli}\ and\ \citenamefont
  {Lousto}(1999{\natexlab{a}})}]{Campanelli99}%
  \BibitemOpen
  \bibfield  {author} {\bibinfo {author} {\bibfnamefont {M.}~\bibnamefont
  {Campanelli}}\ and\ \bibinfo {author} {\bibfnamefont {C.~O.}\ \bibnamefont
  {Lousto}},\ }\href@noop {} {\bibfield  {journal} {\bibinfo  {journal} {Phys.
  Rev.}\ }\textbf {\bibinfo {volume} {D59}},\ \bibinfo {pages} {124022}
  (\bibinfo {year} {1999}{\natexlab{a}})},\ \Eprint
  {https://arxiv.org/abs/gr-qc/9811019} {gr-qc/9811019} \BibitemShut {NoStop}%
%%CITATION = PHRVA,D59,124022;%%
\bibitem [{\citenamefont {Lousto}\ and\ \citenamefont
  {Zlochower}(2007)}]{Lousto:2007mh}%
  \BibitemOpen
  \bibfield  {author} {\bibinfo {author} {\bibfnamefont {C.~O.}\ \bibnamefont
  {Lousto}}\ and\ \bibinfo {author} {\bibfnamefont {Y.}~\bibnamefont
  {Zlochower}},\ }\href@noop {} {\bibfield  {journal} {\bibinfo  {journal}
  {Phys. Rev.}\ }\textbf {\bibinfo {volume} {D76}},\ \bibinfo {pages}
  {041502(R)} (\bibinfo {year} {2007})},\ \Eprint
  {https://arxiv.org/abs/gr-qc/0703061} {gr-qc/0703061} \BibitemShut {NoStop}%
%%CITATION = GR-QC/0703061;%%
\bibitem [{\citenamefont {Nakano}\ \emph {et~al.}(2015)\citenamefont {Nakano},
  \citenamefont {Healy}, \citenamefont {Lousto},\ and\ \citenamefont
  {Zlochower}}]{Nakano:2015pta}%
  \BibitemOpen
  \bibfield  {author} {\bibinfo {author} {\bibfnamefont {H.}~\bibnamefont
  {Nakano}}, \bibinfo {author} {\bibfnamefont {J.}~\bibnamefont {Healy}},
  \bibinfo {author} {\bibfnamefont {C.~O.}\ \bibnamefont {Lousto}},\ and\
  \bibinfo {author} {\bibfnamefont {Y.}~\bibnamefont {Zlochower}},\ }\href
  {https://doi.org/10.1103/PhysRevD.91.104022} {\bibfield  {journal} {\bibinfo
  {journal} {Phys. Rev.}\ }\textbf {\bibinfo {volume} {D91}},\ \bibinfo {pages}
  {104022} (\bibinfo {year} {2015})},\ \Eprint
  {https://arxiv.org/abs/1503.00718} {arXiv:1503.00718 [gr-qc]} \BibitemShut
  {NoStop}%
%%CITATION = ARXIV:1503.00718;%%
\bibitem [{\citenamefont {Healy}\ \emph {et~al.}(2014)\citenamefont {Healy},
  \citenamefont {Lousto},\ and\ \citenamefont {Zlochower}}]{Healy:2014yta}%
  \BibitemOpen
  \bibfield  {author} {\bibinfo {author} {\bibfnamefont {J.}~\bibnamefont
  {Healy}}, \bibinfo {author} {\bibfnamefont {C.~O.}\ \bibnamefont {Lousto}},\
  and\ \bibinfo {author} {\bibfnamefont {Y.}~\bibnamefont {Zlochower}},\ }\href
  {https://doi.org/10.1103/PhysRevD.90.104004} {\bibfield  {journal} {\bibinfo
  {journal} {Phys. Rev.}\ }\textbf {\bibinfo {volume} {D90}},\ \bibinfo {pages}
  {104004} (\bibinfo {year} {2014})},\ \Eprint
  {https://arxiv.org/abs/1406.7295} {arXiv:1406.7295 [gr-qc]} \BibitemShut
  {NoStop}%
%%CITATION = ARXIV:1406.7295;%%
\bibitem [{\citenamefont {Healy}\ and\ \citenamefont
  {Lousto}(2017)}]{Healy:2016lce}%
  \BibitemOpen
  \bibfield  {author} {\bibinfo {author} {\bibfnamefont {J.}~\bibnamefont
  {Healy}}\ and\ \bibinfo {author} {\bibfnamefont {C.~O.}\ \bibnamefont
  {Lousto}},\ }\href {https://doi.org/10.1103/PhysRevD.95.024037} {\bibfield
  {journal} {\bibinfo  {journal} {Phys. Rev.}\ }\textbf {\bibinfo {volume}
  {D95}},\ \bibinfo {pages} {024037} (\bibinfo {year} {2017})},\ \Eprint
  {https://arxiv.org/abs/1610.09713} {arXiv:1610.09713 [gr-qc]} \BibitemShut
  {NoStop}%
%%CITATION = ARXIV:1610.09713;%%
\bibitem [{\citenamefont {Healy}\ \emph
  {et~al.}(2017{\natexlab{a}})\citenamefont {Healy}, \citenamefont {Lousto},\
  and\ \citenamefont {Zlochower}}]{Healy:2017mvh}%
  \BibitemOpen
  \bibfield  {author} {\bibinfo {author} {\bibfnamefont {J.}~\bibnamefont
  {Healy}}, \bibinfo {author} {\bibfnamefont {C.~O.}\ \bibnamefont {Lousto}},\
  and\ \bibinfo {author} {\bibfnamefont {Y.}~\bibnamefont {Zlochower}},\ }\href
  {https://doi.org/10.1103/PhysRevD.96.024031} {\bibfield  {journal} {\bibinfo
  {journal} {Phys. Rev.}\ }\textbf {\bibinfo {volume} {D96}},\ \bibinfo {pages}
  {024031} (\bibinfo {year} {2017}{\natexlab{a}})},\ \Eprint
  {https://arxiv.org/abs/1705.07034} {arXiv:1705.07034 [gr-qc]} \BibitemShut
  {NoStop}%
%%CITATION = ARXIV:1705.07034;%%
\bibitem [{\citenamefont {Zlochower}\ \emph {et~al.}(2017)\citenamefont
  {Zlochower}, \citenamefont {Healy}, \citenamefont {Lousto},\ and\
  \citenamefont {Ruchlin}}]{Zlochower:2017bbg}%
  \BibitemOpen
  \bibfield  {author} {\bibinfo {author} {\bibfnamefont {Y.}~\bibnamefont
  {Zlochower}}, \bibinfo {author} {\bibfnamefont {J.}~\bibnamefont {Healy}},
  \bibinfo {author} {\bibfnamefont {C.~O.}\ \bibnamefont {Lousto}},\ and\
  \bibinfo {author} {\bibfnamefont {I.}~\bibnamefont {Ruchlin}},\ }\href
  {https://doi.org/10.1103/PhysRevD.96.044002} {\bibfield  {journal} {\bibinfo
  {journal} {Phys. Rev.}\ }\textbf {\bibinfo {volume} {D96}},\ \bibinfo {pages}
  {044002} (\bibinfo {year} {2017})},\ \Eprint
  {https://arxiv.org/abs/1706.01980} {arXiv:1706.01980 [gr-qc]} \BibitemShut
  {NoStop}%
%%CITATION = ARXIV:1706.01980;%%
\bibitem [{\citenamefont {Healy}\ \emph
  {et~al.}(2018{\natexlab{b}})\citenamefont {Healy}, \citenamefont {Lousto},
  \citenamefont {Ruchlin},\ and\ \citenamefont {Zlochower}}]{Healy:2017vuz}%
  \BibitemOpen
  \bibfield  {author} {\bibinfo {author} {\bibfnamefont {J.}~\bibnamefont
  {Healy}}, \bibinfo {author} {\bibfnamefont {C.~O.}\ \bibnamefont {Lousto}},
  \bibinfo {author} {\bibfnamefont {I.}~\bibnamefont {Ruchlin}},\ and\ \bibinfo
  {author} {\bibfnamefont {Y.}~\bibnamefont {Zlochower}},\ }\href
  {https://doi.org/10.1103/PhysRevD.97.104026} {\bibfield  {journal} {\bibinfo
  {journal} {Phys. Rev.}\ }\textbf {\bibinfo {volume} {D97}},\ \bibinfo {pages}
  {104026} (\bibinfo {year} {2018}{\natexlab{b}})},\ \Eprint
  {https://arxiv.org/abs/1711.09041} {arXiv:1711.09041 [gr-qc]} \BibitemShut
  {NoStop}%
%%CITATION = ARXIV:1711.09041;%%
\bibitem [{\citenamefont {Healy}\ \emph
  {et~al.}(2017{\natexlab{b}})\citenamefont {Healy}, \citenamefont {Lousto},
  \citenamefont {Zlochower},\ and\ \citenamefont {Campanelli}}]{Healy:2017psd}%
  \BibitemOpen
  \bibfield  {author} {\bibinfo {author} {\bibfnamefont {J.}~\bibnamefont
  {Healy}}, \bibinfo {author} {\bibfnamefont {C.~O.}\ \bibnamefont {Lousto}},
  \bibinfo {author} {\bibfnamefont {Y.}~\bibnamefont {Zlochower}},\ and\
  \bibinfo {author} {\bibfnamefont {M.}~\bibnamefont {Campanelli}},\ }\href
  {https://doi.org/10.1088/1361-6382/aa91b1} {\bibfield  {journal} {\bibinfo
  {journal} {Class. Quant. Grav.}\ }\textbf {\bibinfo {volume} {34}},\ \bibinfo
  {pages} {224001} (\bibinfo {year} {2017}{\natexlab{b}})},\ \Eprint
  {https://arxiv.org/abs/1703.03423} {arXiv:1703.03423 [gr-qc]} \BibitemShut
  {NoStop}%
%%CITATION = ARXIV:1703.03423;%%
\bibitem [{\citenamefont {Campanelli}\ and\ \citenamefont
  {Lousto}(1999{\natexlab{b}})}]{Campanelli:1998jv}%
  \BibitemOpen
  \bibfield  {author} {\bibinfo {author} {\bibfnamefont {M.}~\bibnamefont
  {Campanelli}}\ and\ \bibinfo {author} {\bibfnamefont {C.~O.}\ \bibnamefont
  {Lousto}},\ }\href {https://doi.org/10.1103/PhysRevD.59.124022} {\bibfield
  {journal} {\bibinfo  {journal} {Phys. Rev.}\ }\textbf {\bibinfo {volume}
  {D59}},\ \bibinfo {pages} {124022} (\bibinfo {year} {1999}{\natexlab{b}})},\
  \Eprint {https://arxiv.org/abs/gr-qc/9811019} {arXiv:gr-qc/9811019 [gr-qc]}
  \BibitemShut {NoStop}%
\bibitem [{\citenamefont {Gundlach}\ and\ \citenamefont
  {Martin-Garcia}(2006)}]{Gundlach:2006tw}%
  \BibitemOpen
  \bibfield  {author} {\bibinfo {author} {\bibfnamefont {C.}~\bibnamefont
  {Gundlach}}\ and\ \bibinfo {author} {\bibfnamefont {J.~M.}\ \bibnamefont
  {Martin-Garcia}},\ }\href@noop {} {\bibfield  {journal} {\bibinfo  {journal}
  {Phys. Rev.}\ }\textbf {\bibinfo {volume} {D74}},\ \bibinfo {pages} {024016}
  (\bibinfo {year} {2006})},\ \Eprint {https://arxiv.org/abs/gr-qc/0604035}
  {gr-qc/0604035} \BibitemShut {NoStop}%
%%CITATION = GR-QC 0604035;%%
\bibitem [{\citenamefont {M{\"u}ller}\ and\ \citenamefont
  {Br{\"u}gmann}(2010)}]{Mueller:2009jx}%
  \BibitemOpen
  \bibfield  {author} {\bibinfo {author} {\bibfnamefont {D.}~\bibnamefont
  {M{\"u}ller}}\ and\ \bibinfo {author} {\bibfnamefont {B.}~\bibnamefont
  {Br{\"u}gmann}},\ }\href@noop {} {\bibfield  {journal} {\bibinfo  {journal}
  {Class. Quant. Grav.}\ }\textbf {\bibinfo {volume} {27}},\ \bibinfo {pages}
  {114008} (\bibinfo {year} {2010})},\ \Eprint
  {https://arxiv.org/abs/0912.3125} {arXiv:0912.3125 [gr-qc]} \BibitemShut
  {NoStop}%
%%CITATION = 0912.3125;%%
\bibitem [{\citenamefont {M{\"u}ller}\ \emph
  {et~al.}(2010{\natexlab{a}})\citenamefont {M{\"u}ller}, \citenamefont
  {Grigsby},\ and\ \citenamefont {Br{\"u}gmann}}]{Mueller:2010bu}%
  \BibitemOpen
  \bibfield  {author} {\bibinfo {author} {\bibfnamefont {D.}~\bibnamefont
  {M{\"u}ller}}, \bibinfo {author} {\bibfnamefont {J.}~\bibnamefont
  {Grigsby}},\ and\ \bibinfo {author} {\bibfnamefont {B.}~\bibnamefont
  {Br{\"u}gmann}},\ }\href {https://doi.org/10.1103/PhysRevD.82.064004}
  {\bibfield  {journal} {\bibinfo  {journal} {Phys. Rev.}\ }\textbf {\bibinfo
  {volume} {D82}},\ \bibinfo {pages} {064004} (\bibinfo {year}
  {2010}{\natexlab{a}})},\ \Eprint {https://arxiv.org/abs/1003.4681}
  {arXiv:1003.4681 [gr-qc]} \BibitemShut {NoStop}%
%%CITATION = 1003.4681;%%
\bibitem [{\citenamefont {Schnetter}(2010)}]{Schnetter:2010cz}%
  \BibitemOpen
  \bibfield  {author} {\bibinfo {author} {\bibfnamefont {E.}~\bibnamefont
  {Schnetter}},\ }\href {https://doi.org/10.1088/0264-9381/27/16/167001}
  {\bibfield  {journal} {\bibinfo  {journal} {Class. Quant. Grav.}\ }\textbf
  {\bibinfo {volume} {27}},\ \bibinfo {pages} {167001} (\bibinfo {year}
  {2010})},\ \Eprint {https://arxiv.org/abs/1003.0859} {arXiv:1003.0859
  [gr-qc]} \BibitemShut {NoStop}%
%%CITATION = 1003.0859;%%
\bibitem [{\citenamefont {Alic}\ \emph {et~al.}(2010)\citenamefont {Alic},
  \citenamefont {Rezzolla}, \citenamefont {Hinder},\ and\ \citenamefont
  {Mosta}}]{Alic:2010wu}%
  \BibitemOpen
  \bibfield  {author} {\bibinfo {author} {\bibfnamefont {D.}~\bibnamefont
  {Alic}}, \bibinfo {author} {\bibfnamefont {L.}~\bibnamefont {Rezzolla}},
  \bibinfo {author} {\bibfnamefont {I.}~\bibnamefont {Hinder}},\ and\ \bibinfo
  {author} {\bibfnamefont {P.}~\bibnamefont {Mosta}},\ }\href
  {https://doi.org/10.1088/0264-9381/27/24/245023} {\bibfield  {journal}
  {\bibinfo  {journal} {Class. Quant. Grav.}\ }\textbf {\bibinfo {volume}
  {27}},\ \bibinfo {pages} {245023} (\bibinfo {year} {2010})},\ \Eprint
  {https://arxiv.org/abs/1008.2212} {arXiv:1008.2212 [gr-qc]} \BibitemShut
  {NoStop}%
%%CITATION = 1008.2212;%%
\bibitem [{\citenamefont {Lousto}\ and\ \citenamefont
  {Zlochower}(2011)}]{Lousto:2010ut}%
  \BibitemOpen
  \bibfield  {author} {\bibinfo {author} {\bibfnamefont {C.~O.}\ \bibnamefont
  {Lousto}}\ and\ \bibinfo {author} {\bibfnamefont {Y.}~\bibnamefont
  {Zlochower}},\ }\href {https://doi.org/10.1103/PhysRevLett.106.041101}
  {\bibfield  {journal} {\bibinfo  {journal} {Phys. Rev. Lett.}\ }\textbf
  {\bibinfo {volume} {106}},\ \bibinfo {pages} {041101} (\bibinfo {year}
  {2011})},\ \Eprint {https://arxiv.org/abs/1009.0292} {arXiv:1009.0292
  [gr-qc]} \BibitemShut {NoStop}%
%%CITATION = 1009.0292;%%
\bibitem [{\citenamefont {Bernard}\ \emph {et~al.}(2018)\citenamefont
  {Bernard}, \citenamefont {Blanchet}, \citenamefont {Faye},\ and\
  \citenamefont {Marchand}}]{Bernard:2017ktp}%
  \BibitemOpen
  \bibfield  {author} {\bibinfo {author} {\bibfnamefont {L.}~\bibnamefont
  {Bernard}}, \bibinfo {author} {\bibfnamefont {L.}~\bibnamefont {Blanchet}},
  \bibinfo {author} {\bibfnamefont {G.}~\bibnamefont {Faye}},\ and\ \bibinfo
  {author} {\bibfnamefont {T.}~\bibnamefont {Marchand}},\ }\href
  {https://doi.org/10.1103/PhysRevD.97.044037} {\bibfield  {journal} {\bibinfo
  {journal} {Phys. Rev.}\ }\textbf {\bibinfo {volume} {D97}},\ \bibinfo {pages}
  {044037} (\bibinfo {year} {2018})},\ \Eprint
  {https://arxiv.org/abs/1711.00283} {arXiv:1711.00283 [gr-qc]} \BibitemShut
  {NoStop}%
%%CITATION = ARXIV:1711.00283;%%
\bibitem [{\citenamefont {Marronetti}\ \emph {et~al.}(2008)\citenamefont
  {Marronetti}, \citenamefont {Tichy}, \citenamefont {Br{\"u}gmann},
  \citenamefont {Gonzalez},\ and\ \citenamefont
  {Sperhake}}]{Marronetti:2007wz}%
  \BibitemOpen
  \bibfield  {author} {\bibinfo {author} {\bibfnamefont {P.}~\bibnamefont
  {Marronetti}}, \bibinfo {author} {\bibfnamefont {W.}~\bibnamefont {Tichy}},
  \bibinfo {author} {\bibfnamefont {B.}~\bibnamefont {Br{\"u}gmann}}, \bibinfo
  {author} {\bibfnamefont {J.}~\bibnamefont {Gonzalez}},\ and\ \bibinfo
  {author} {\bibfnamefont {U.}~\bibnamefont {Sperhake}},\ }\href
  {https://doi.org/10.1103/PhysRevD.77.064010} {\bibfield  {journal} {\bibinfo
  {journal} {Phys. Rev.}\ }\textbf {\bibinfo {volume} {D77}},\ \bibinfo {pages}
  {064010} (\bibinfo {year} {2008})},\ \Eprint
  {https://arxiv.org/abs/0709.2160} {arXiv:0709.2160 [gr-qc]} \BibitemShut
  {NoStop}%
%%CITATION = 0709.2160;%%
\bibitem [{\citenamefont {Lousto}\ \emph
  {et~al.}(2010{\natexlab{b}})\citenamefont {Lousto}, \citenamefont {Nakano},
  \citenamefont {Zlochower},\ and\ \citenamefont {Campanelli}}]{Lousto:2010tb}%
  \BibitemOpen
  \bibfield  {author} {\bibinfo {author} {\bibfnamefont {C.~O.}\ \bibnamefont
  {Lousto}}, \bibinfo {author} {\bibfnamefont {H.}~\bibnamefont {Nakano}},
  \bibinfo {author} {\bibfnamefont {Y.}~\bibnamefont {Zlochower}},\ and\
  \bibinfo {author} {\bibfnamefont {M.}~\bibnamefont {Campanelli}},\
  }\href@noop {} {\bibfield  {journal} {\bibinfo  {journal} {Phys. Rev. Lett.}\
  }\textbf {\bibinfo {volume} {104}},\ \bibinfo {pages} {211101} (\bibinfo
  {year} {2010}{\natexlab{b}})},\ \Eprint {https://arxiv.org/abs/1001.2316}
  {arXiv:1001.2316 [gr-qc]} \BibitemShut {NoStop}%
%%CITATION = 1001.2316;%%
\bibitem [{\citenamefont {M{\"u}ller}\ \emph
  {et~al.}(2010{\natexlab{b}})\citenamefont {M{\"u}ller}, \citenamefont
  {Grigsby},\ and\ \citenamefont {Br{\"u}gmann}}]{Muller:2010zze}%
  \BibitemOpen
  \bibfield  {author} {\bibinfo {author} {\bibfnamefont {D.}~\bibnamefont
  {M{\"u}ller}}, \bibinfo {author} {\bibfnamefont {J.}~\bibnamefont
  {Grigsby}},\ and\ \bibinfo {author} {\bibfnamefont {B.}~\bibnamefont
  {Br{\"u}gmann}},\ }\href {https://doi.org/10.1103/PhysRevD.82.064004}
  {\bibfield  {journal} {\bibinfo  {journal} {Phys. Rev.}\ }\textbf {\bibinfo
  {volume} {D82}},\ \bibinfo {pages} {064004} (\bibinfo {year}
  {2010}{\natexlab{b}})},\ \Eprint {https://arxiv.org/abs/1003.4681}
  {arXiv:1003.4681 [gr-qc]} \BibitemShut {NoStop}%
%%CITATION = 1003.4681;%%
\end{thebibliography}%

\end{document}